\documentclass{emulateapj}
\newcommand{\hst}{{\emph{HST}}}

\newcommand{\iras}{{\emph{IRAS}}}
\newcommand{\spitzer}{{\emph{Spitzer}}}

\newcommand{\kms}{\mbox{\,km\,s$^{-1}$}}

\newcommand{\lumin}{\mbox{\,ergs~s$^{-1}$}}

\newcommand{\HI}{\ion{H}{1}}

\newcommand{\altcite}{\citealt}

\newcommand{\total}{46}

\newcommand{\microns}{\micron}

\newcommand{\lsun}{$L_{\odot}$}
\newcommand{\msun}{$M_{\odot}$}

\newcommand{\lfour}{$L_{\rm 4.5}$}
\newcommand{\ltwofour}{$L_{\rm 24}$}
\newcommand{\leight}{$L_{\rm8}$}
\newcommand{\alphairac}{$\alpha_{\rm IRAC}$}

\newcommand{\alphathree}{$\alpha_{8,24}$}
\newcommand{\frachd}{$f_{\rm HD}$}
\newcommand{\fracd}{$f_{\rm 24D}$}

\begin{document}
 
 
\shortauthors{Gallagher et al.}
\shorttitle{Mid-Infrared Activity in Compact Galaxy Groups}

\title{The Revealing Dust: Mid-Infrared Activity in Hickson Compact
Group Galaxy Nuclei}

\author{S.\ C. Gallagher,\altaffilmark{1}
K.\ E. Johnson,\altaffilmark{2,3}
A.\ E. Hornschemeier,\altaffilmark{4}
J.\ C. Charlton,\altaffilmark{5}
\& J.\ E. Hibbard\altaffilmark{6}
}

\altaffiltext{1}{Department of Physics and Astronomy, University of
  California -- Los Angeles, Los Angeles CA, 90095--1547, USA; {\em sgall@astro.ucla.edu}}
\altaffiltext{2}{University of Virginia, Charlottesville, VA, USA}
\altaffiltext{3}{National Radio Astronomical Observatory,
  Charlottesville, VA, USA}
\altaffiltext{4}{NASA's Goddard Space Flight Center, Greenbelt, MD, USA}
\altaffiltext{5}{Department of Astronomy and Astrophysics,
  Pennsylvania State University, University Park, PA
  16801, USA}
       
\begin{abstract}
We present a sample of \total\ galaxy nuclei from 12 nearby
($z<4500$\kms) Hickson Compact Groups (HCGs) with a complete suite of
1--24\micron\ 2MASS+\spitzer\ nuclear photometry.  For all objects in
the sample, blue emission from stellar photospheres dominates in the
near-infrared through the 3.6\micron\ IRAC band.  Twenty-five of
\total\ (54\%) galaxy nuclei show red, mid-infrared continua
characteristic of hot dust powered by ongoing star formation and/or
accretion onto a central black hole. We introduce 
\alphairac, the spectral index of a power-law fit to the
4.5--8.0\micron\ IRAC data, and demonstrate that it cleanly separates
the mid-infrared active and non-active HCG nuclei.  This
parameter is more powerful for identifying low to moderate-luminosity
mid-infrared activity than other measures which include data at
rest-frame $\lambda<3.6$\micron\ that may be dominated by stellar
photospheric emission.  While the HCG galaxies clearly have a bimodal
distribution in this parameter space, a comparison sample from the
\spitzer\ Nearby Galaxy Survey (SINGS) matched in $J$-band total
galaxy luminosity is continuously distributed.  A second diagnostic,
the fraction of 24\micron\ emission in excess of that expected from
quiescent galaxies, \fracd, reveals an additional three nuclei to be
active at 24\micron.  Comparing these two mid-infrared diagnostics of
nuclear activity to optical spectroscopic identifications from the
literature reveals some discrepancies, and we discuss the challenges
of distinguishing the source of ionizing radiation in these and other
lower luminosity systems.  We find a significant correlation between
the fraction of mid-infrared active galaxies and the total \HI\ mass
in a group, and investigate possible interpretations of these results
in light of galaxy evolution in the highly interactive system of a compact
group environment.
\end{abstract}


\section{Introduction}
\label{sec:intro}

The first galaxies and their environments differed substantially from
those locally, often involving multiple interactions
as seen in the \hst\ Ultra Deep Field \citep[e.g.,][]{malhotra05}.
Compared to all other nearby environments, present-day compact galaxy
groups most closely reproduce the interaction environment of the early
Universe ($z\sim4$) when galaxies assembled through hierarchical formation
\citep[e.g.,][]{baron}, and galaxy groups combined to form
proto-clusters (in dense regions; e.g., \altcite{rudick+06}) or massive
ellipticals (in the field; \altcite{white}).

Because of their high space densities (with comparable surface
densities to the centers of rich galaxy clusters; e.g.,
\altcite{rubin+91}) and low velocity dispersions ($\sigma_{\rm
v}\sim10^2$~\kms), compact groups of galaxies are ideal environments
for studying the mechanisms of interaction-induced star formation and
nuclear activity.  From optical spectroscopic surveys, Hickson Compact
Groups (HCGs) are known to host a population of galaxies with
emission-line nuclear spectra characteristic of star-formation and/or
active galactic nuclei (AGNs).  Based on the optical spectroscopic
survey of \citet{coziol98a,coziol98b}, the AGN fraction in HCGs is
found to be $\sim40\%$, perhaps consistent with the $43\%$ nuclear activity
level found for nearby $M_V<-19$ galaxies (with greater detection sensitivity;
\altcite{HoEtal97a}) and significantly higher than the $\sim1\%$ AGN
fraction identified optically in cluster galaxies (with $M_{V}<-21$;
\altcite{dressler85}).  Further, HCG AGNs (including low-luminosity
AGNs, hereafter LLAGNs; HCGs host no known Seyfert~1-luminosity AGNs)
are preferentially found in optically luminous, early-type galaxies
with little or no ongoing star formation in the cores of evolved
groups.  Similarly, many galaxies in clusters host
LLAGNs in the local Universe \citep[e.g.,][]{martini07}, in particular
brightest cluster galaxies \citep{best07}.

Typically, the most active star-forming galaxies are late-type spirals
in the outskirts of groups. \citet{coziol98b} interpreted relative
distributions of star-forming and AGN-hosting galaxies as indicating a
clear evolutionary scenario whereby group cores are mature collapsed
systems in which the high galaxy densities led to increased
gravitational interactions and hence more rapid exhaustion of gas
reservoirs in the past through star formation.  (The radiatively weak
AGNs in the evolved group cores require only a small amount of gas for
fueling.)  A similar study by
\citet{shimada00} found that the fraction of emission-line HCG
galaxies was comparable to the field, and conversely concluded that
the HCG environment does {\it not} trigger either star formation or
AGN activity.

The initial expectation that the interactions evident in the compact
group environment would naturally lead to markedly enhanced levels of star
formation compared with the field has not been satisfied, and a
coherent understanding of the history of gas and cold dust in compact
groups has proven elusive.  An analysis of the CO content in HCG
galaxies found them to be similar to those in the field, in loose
groups, and in other environments; a notable exception is the
$\sim20\%$ of HCG spirals that are CO deficient.  This suggests less,
not more, star formation in HCGs compared to other environments
\citep{verdes98}. At the same time, the
detection of a few HCG elliptical and S0 galaxies in both CO and the
far-infrared (unlike typical galaxies of these types) suggests that
tidal interactions are influencing galaxy evolution to some
extent. While the far-infrared power is similar to comparison samples,
the ratios of 25 to 100\micron\ \iras\ fluxes implies a greater number
of intense, nuclear starbursts in HCG galaxies \citep{verdes98}.

Clearly, a robust and consistent understanding of the impact of the
compact group environment on galaxy properties has not yet emerged.
One possible difficulty to date is the predominant use of optical
emission-line studies to identify activity --- both star-forming and
accretion-dominated.  Ground-based studies of bright galaxies can
easily obscure low-contrast emission lines through their dilution by a
strong stellar contribution \citep[e.g.,][]{HoEtal97c}; intrinsic
absorption can also mask spectroscopic signatures.  The clear
discrepancy between the $\sim1\%$ AGN fraction in galaxy clusters from
optical spectroscopic surveys \citep[e.g.,][]{dressler85} compared to
the larger fraction ($\sim5\%$ for luminous galaxies) revealed by
X-ray observations \citep[e.g.,][]{martini06} is illustrative of this
problem. As highlighted by \citet{martini07}, mismatched
sample selection and detection techniques can create apparent (and
false) discrepancies in AGN fractions between environments.
In this paper, we take an alternate approach, focusing on the
mid-infrared spectral energy distributions (SEDs) of individual
galaxy nuclei to clearly identify the thermal, hot dust continua that
signify neutral gas heated by ionizing photons from either young stars
or an AGN.  The clear discrepancy between a blue, quiescent galaxy SED
where the mid-infrared is dominated by the Raleigh-Jeans tail of
stellar photospheres and the red, mid-infrared SED of warm to hot dust
emission offers promise for reducing the ambiguity of previous compact
group studies.

\citet[][hereafter J07]{kelsey07} have presented the first results
from a Cycle~1 \spitzer\ IRAC (3.6--8.0\micron) and MIPS (24\micron)
imaging survey of \total\ galaxies in 12 nearby HCGs.  In brief, this
work revealed trends between the evolutionary states of compact groups
(determined from their dynamical and \HI\ masses), and their
mid-infrared colors and luminosities.  Galaxies in relatively gas-rich
groups tend to have colors most indicative of star formation and AGN
activity, and galaxies in gas-poor groups predominantly exhibit a
narrow range of mid-infrared colors that are consistent with the light
from quiescent stellar populations.  The galaxies in this sample of 12
compact groups also occupy infrared color space in a distinctly
different way than the population of galaxies in the {\it Spitzer}
First Look Survey (FLS; e.g., \altcite{lacy04}), notably exhibiting a
``gap'' in their color distribution not found in the FLS sample.  All
of these results suggest that the environment of a compact group is
intimately connected to the mid-infrared activity of the member
galaxies.  We present additional analysis of the nuclei of these
galaxies, focusing in particular on developing diagnostics for
identifying quantitative measures of mid-infrared activity with the
ultimate goal of exploring the nature of mid-infrared emission, i.e.,
quiescent, star-forming, and/or accretion-powered.  We also explore
the connections between mid-infrared nuclear properties and other features of
individual galaxies including morphology, optical spectral type, and
the host group's evolutionary stage.  Because of their proximity
($v<4500$\kms), we can spatially extract nuclear photometry for the
HCG galaxies to reduce contamination from the extended galaxy that is
inevitable in surveys of more distant galaxies.

Throughout we assume a $\Lambda$-CDM cosmology with $\Omega_{\rm
M}=0.3$, $\Omega_{\Lambda}=0.7$, and $H_0=70$\kms\,Mpc$^{-1}$
\citep[e.g.,][]{sperg07}.

\section{Data Reduction and Analysis}

\subsection{Sample}

Target compact galaxy groups for the \spitzer\ program were selected
on the basis of distance, angular extent, and membership (at least
three members with accordant redshifts, i.e., within 1000\kms\ of the
group mean).  All HCGs were considered at redshifts $<4500$\kms, and
the sample includes all of the nearest ones with only a few exceptions
--- three nearby groups are too extended to realistically cover with
small-field imagers.  The morphologies of the individual galaxies in
the groups include tidal arms and tails, polar rings, ellipticals with
boxy isophotes and central disks, and disturbed features; typical
spiral and elliptical galaxies are also present.  In
Table~\ref{tab:groups}, we list the groups, average recession
velocities, membership by morphology, mass in neutral hydrogen
($M_{\rm HI}$), a qualitative description of the evolutionary stage of
each group based on the scenario proposed by \citet{verdes}, and the
\HI\ type from J07. Groups are divided into \HI\ types of I (rich; 
$\log(M_{\rm HI})/\log(M_{\rm dyn})>0.9$), II (intermediate; $\log(M_{\rm
HI})/\log(M_{\rm dyn})=0.8$--0.9), and III (poor; $\log(M_{\rm
HI})/\log(M_{\rm dyn})<0.8$), where $M_{\rm dyn}$ is the 
group dynamical mass (J07 and references therein).

\subsection{Nuclear Photometry}

The observations and image reduction are described in detail in J07,
where the extended source photometry is presented.  For this project,
we focus on the nuclear regions in particular.  The nuclear flux
density is obtained using a circular aperture centered on the nucleus
(with some exceptions; see below) with a 7\arcsec\ radius which
corresponds to 1.1--1.8~kpc for our groups.  Determining the absolute
background in these crowded environments is not straightforward.  We
elected to use a uniform method in which background is taken from an
annulus with inner and outer radii of 50\arcsec\ and 60\arcsec,
respectively. This annulus is well outside the 3$\sigma$ contour
levels of nearly all of the galaxies in the sample with the exception
of galaxies in close projected proximity with overlapping contours.
However, since the mode of the annulus is used to determine the
background level in all cases, partial contamination from a nearby
source will not affect the measured background values. For the
purposes of comparing photometry at different wavelengths, the IRAC
and MIPS images were cross-convolved with their respective point
response functions (PRFs) to obtain images with matched spatial
resolution.  No aperture corrections have been applied. For
reference, a 7\arcsec\ radius aperture corresponds to $\sim70\%$ of
the encircled energy of the 24\micron\ PRF.

In most cases, the nuclear aperture is centered on the location of the
central flux peak of a galaxy.  However, in the case of
disrupted/merging/irregular galaxies, some discretion is required.
For example, for the galaxies 31a, c, and e in HCG~31, the
infrared peak apparently at the galaxy collision interface is
designated as the nucleus of the main interacting pair.

The relative errors listed for the nuclear regions in Table~\ref{tab:fluxes}
are conservatively set at 10\%.  The main source of relative
uncertainty between the bands is the agreement of the relative
astrometry; this translates into how well-centered the nuclear region
is in each band.  Empirically, this is found to be $<5\%$, and usually
2--3\%.  The formal uncertainties derived from the image statistics
are negligible by comparison.

\begin{figure*}
\figurenum{1}
\plotone{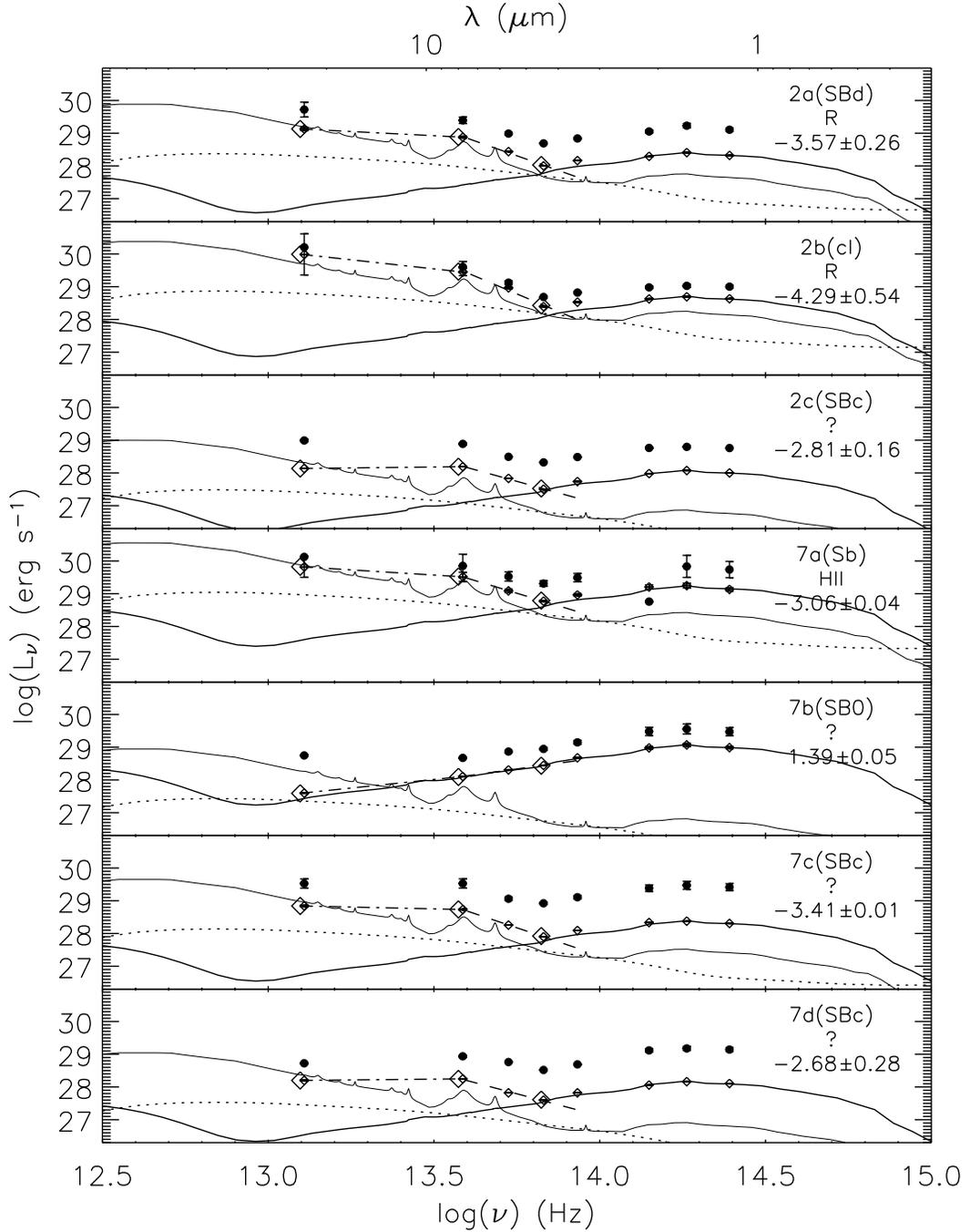}
\caption{SEDs for the first seven HCG galaxies in rest-frame units. 
Nuclear data (small open diamonds) are listed in
Table~\ref{tab:fluxes}; the total galaxy photometry (filled circles)
is from J07. For reference, a composite AGN SED (dotted curve;
\altcite{ric+06}) and the model M82 SED (thin solid curve;
\altcite{grasil_ref}) have been overplotted on each panel; both were
normalized to the 4.5\micron\ nuclear luminosity minus the stellar
contribution from an elliptical galaxy template (thick solid curve;
\altcite{grasil_ref}).  
The power-law fit ($L_{\nu}\propto\nu^{\alpha_{\rm IRAC}}$) to the
4.5--8.0\micron\ photometry is shown as the dashed line, and the
dot-dashed line indicates the \alphathree\ power-law model.  The
values (from left to right) for \ltwofour, \leight, and \lfour\ are indicated with
large open diamonds.  Objects are labeled with HCG name (morphology),
optical spectral type (where known) of the nucleus, and \alphairac\ (errors are
1$\sigma$).}
\label{fig:sed}
\end{figure*}

This photometry procedure is different than the method used for the
entire galaxies presented in J07, which used isocontour levels.  In
some cases, the flux density of the entire galaxy $\times0.7$ (to
roughly account for an aperture correction) is actually less than the
`nuclear' region, and this is because the isocontour levels used for
the extended galaxy photometry fell within the 7\arcsec\ radius used
for the nuclear regions.  However, frequently the aperture-corrected
flux of the nuclear region is essentially equal to that of the galaxy;
this indicates that at this angular resolution the galaxies are
pointlike at 24\micron.  The nuclear photometry from the near-infrared
through 24\micron\ is presented in Table~\ref{tab:fluxes}.  SEDs for
each galaxy nucleus are plotted in rest-frame units in
Figure~\ref{fig:sed}, along with the morphological type and optical
spectroscopic classification (as available).  Also overplotted is the
total galaxy photometry and three SED templates, that of an elliptical
galaxy \citep{grasil_ref}, a Type~1 (i.e., broad emission-line) AGN \citep{ric+06}, and the local
starburst galaxy M82 \citep{grasil_ref}.  These three templates
encompass the range of expected mid-infrared spectral shapes for
galaxies in the local Universe.  In general, the shape of the SEDs is
comparable for the total galaxy and nuclear photometry.

\subsection{Mid-Infrared Diagnostics}
\label{sec:params}

To investigate nuclear activity in HCGs further, we calculate several
values from the photometry to categorize the galaxies.  First, we
measure the luminosities at several wavelengths chosen to characterize
the contribution from different emission mechanisms.  The $J$-band
luminosity, $L_J$, is a measure of the luminosity in stars of a
galaxy.  Unlike in previous studies \citep[e.g.,][]{verdes98} that
used a blue luminosity, $L_B$, ongoing star-formation does not
significantly alter $L_J$. The luminosity at 4.5\microns, \lfour,
measures the stellar continuum plus any contribution from hot
($\gtrsim600$~K) dust. This bandpass has the advantage of containing
no significant polycyclic aromatic hydrocarbon (PAH) features.  However, the
stellar contribution can still be quite significant, and so we also
calculate the fraction of 4.5\micron\ emission from hot dust: \frachd
$= 1-L_{\rm 4.5,\ast}$, where $L_{\rm 4.5,
\ast}$ is the light from stars expected from the elliptical
galaxy template of \citet{grasil_ref} normalized to the $J$ and
$H$-band photometry.  Though not all galaxies in the sample are
ellipticals, we use this template to represent the stellar population,
as templates from spirals and irregulars will include warm dust
emission.  The \citet{grasil_ref} elliptical template contains a
contribution from cooler dust attributed to galactic cirrus, but this
component is not significant for $\lambda < 25$\micron.
Additional fiducial luminosities are measured at 8 and 24\microns.
The former includes PAH emission (if present), and the latter is
sensitive to cooler dust in a frequency regime near the minimum of the
elliptical galaxy template.  The value, \fracd, the fraction of
24\micron\ luminosity not attributed to starlight, is calculated in
the same manner as
\frachd.

Finally, we also calculate the spectral index, \alphairac, for a
power-law model fit to the 4.5--8.0\micron\ data,
$L_{\nu}\propto\nu^{\alpha_{\rm IRAC}}$.  The 3.6\micron\ photometry is
not included because of the dominant stellar contribution.  In
addition, an 8 to 24\micron\ spectral index,
\alphathree=0.477\,$\log(L_8/L_{24})$, can reveal cooler dust or dust
emission diluted by a strong stellar continuum at shorter wavelengths.
All of these values are listed in Table~\ref{tab:props}.

\subsection{Comparison Galaxy Sample}

\begin{figure*}
\plottwo{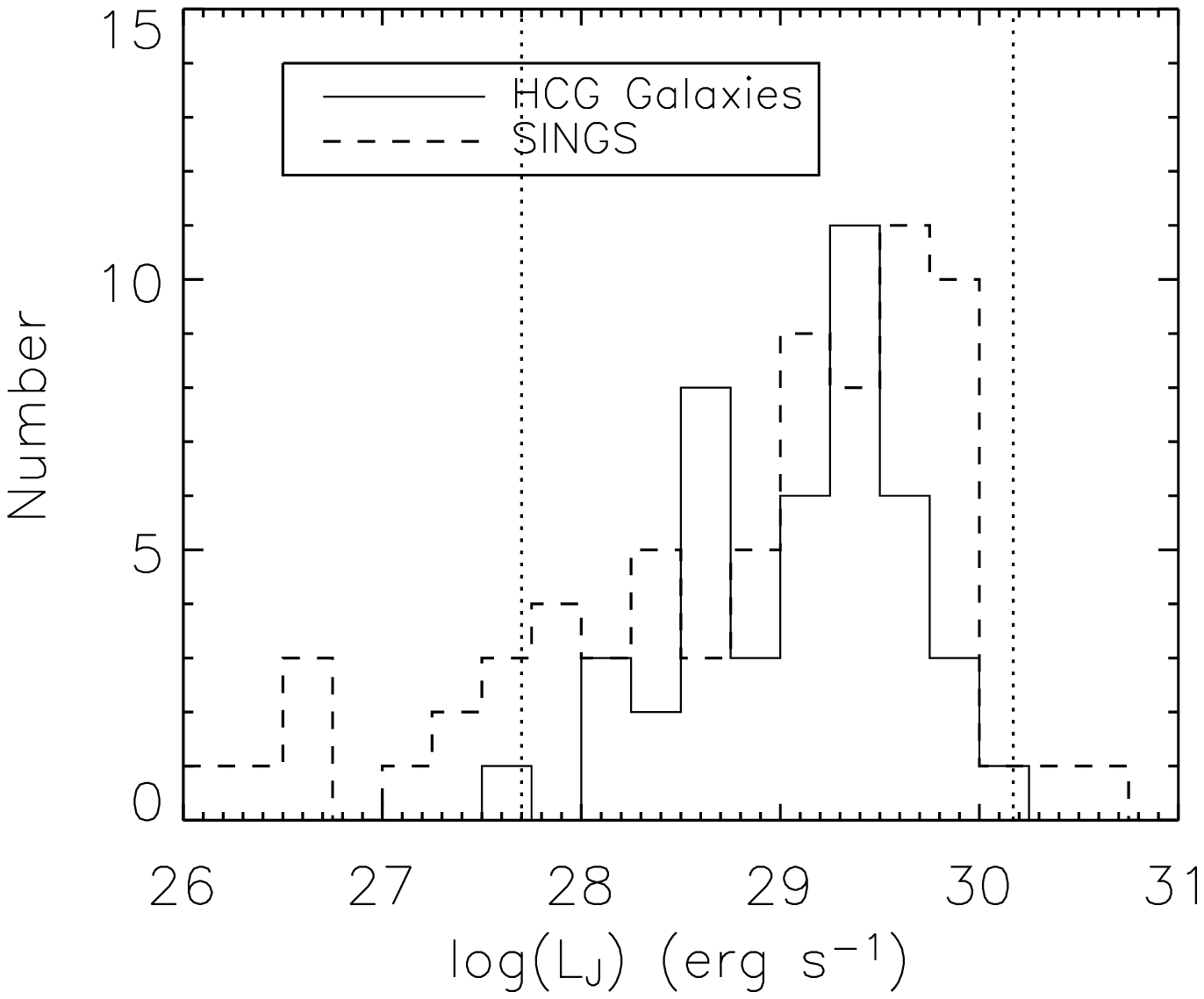}{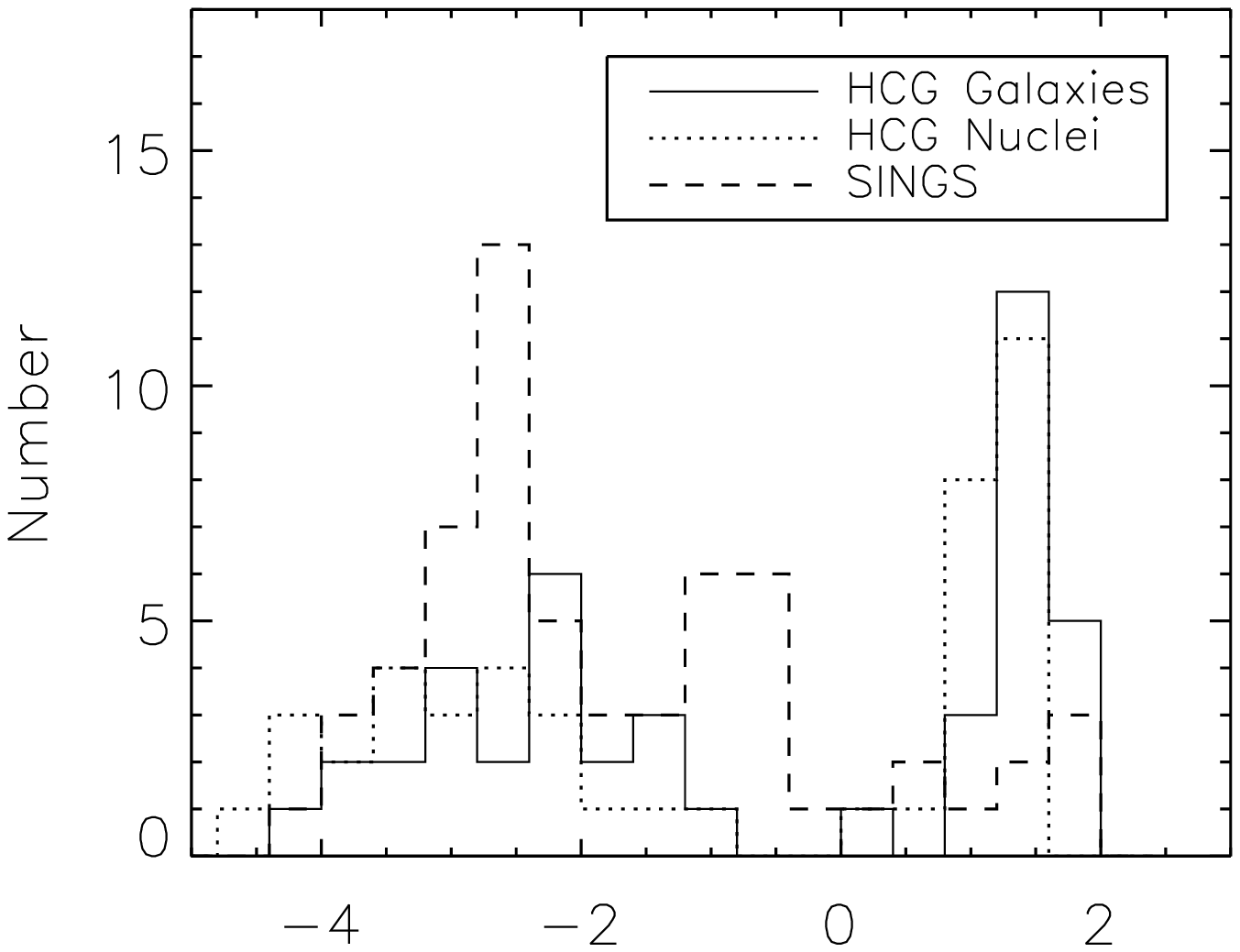}
\caption{Histograms comparing the distribution of integrated galaxy $J$-band luminosity
({\em left panel}) and \alphairac\ ({\em right panel}) for the HCG (solid)
and SINGS comparison galaxy (dashed; \altcite{dale+07}) samples.  Only
the 61 SINGS galaxies between the vertical dotted lines in the left
panel are plotted in
subsequent figures. The \alphairac\
distributions of the two samples are clearly discrepant, with the HCG
galaxies (and nuclei; dotted) showing a deficit for values of \alphairac$\sim-0.5$.  
\label{fig:hists}
}
\end{figure*}

In order to evaluate the role (if any) of the HCG environment on
mid-infrared activity, a comparison galaxy sample is required.  For
this purpose, we use the \spitzer\ Nearby Galaxy Sample (SINGS)
photometry presented in \citet{dale+07}.  This is the largest sample
of comparable \spitzer\ data available in the literature for galaxies
in the local Universe with known distances.  The SINGS sample was
chosen to be diverse, covering a large parameter space of optical
luminosities, masses, metallicities, star-formation rates, and
morphologies.  In an attempt to identify a reasonable comparison
sample of galaxies within SINGS, we have filtered the SINGS sample to
match the values of $L_J$(gal)\footnote{Hereafter, to indicate
integrated galaxy as opposed to nuclear properties for the HCGs,
`(gal)' will be appended to parameter names.} (from J07 for the entire
galaxy, not just the nucleus) found in our sample, $\log(L_J{\rm
(gal)}) = 27.70$--30.17 \lumin~Hz$^{-1}$.  This leaves 61 out of the
total SINGS sample of 75.  The majority of those filtered from the
SINGS sample are low-luminosity dwarfs; the HCG sample comprises
primarily 3--5 bright galaxies per group. The distributions of
$L_J$(gal) for the HCG and SINGS samples are plotted in
Figure~\ref{fig:hists}a.  For all of the SINGS galaxies, we use the
photometry in Tables~2 and 3 of \citet{dale+07} and calculate the same
parameters as for the HCG galaxies detailed in \S\ref{sec:params}.
The extended source photometry for the HCG galaxies includes aperture
corrections to correct for scattered light in the IRAC bands following
Eq.~2 of \citet{dale+07}, and therefore the two datasets are directly
comparable (see J07 for further details).

\section{Results and Discussion}
\label{sec:disc}

From a visual examination of the near through mid-infrared SEDs
(Fig.~1), the close correspondence of the $J$-band through 3.6\micron\
photometric data points with an elliptical galaxy template indicates
that stellar emission dominates in this wavelength regime in every
galaxy.  To account for this, we normalize a 13~Gyr elliptical galaxy
template \citep{grasil_ref} to the $J$ and $H$ band flux densities.
With this simple normalization, the elliptical galaxy SED predicts the
3.6--8.0\micron\ flux densities for 21 of the \total\ (46\%) galaxies
remarkably well.  These include the entire membership of the
evolved and X-ray-bright HCGs 42 and 62.  The remaining 25 galaxies (counting
HCG~31ace as 3 galaxies), have mid-infrared continua that rise to
longer wavelengths.  This clear qualitative difference is illustrated
quantitatively with the histogram of
\alphairac\ values shown in Figure~\ref{fig:hists}b.  No HCG galaxy nuclei
inhabit the parameter space between
\alphairac\ values of $-0.95$ and 0.25; for the integrated HCG galaxy
SEDs, the distribution is comparable, with none found in the range of
\alphairac(gal) of $-0.95$ to 0.12. 
This is in stark contrast to the distribution of \alphairac\ values of
the SINGS galaxies (plotted in the same figure) that shows a
significant fraction within the HCG gap.  We also point out a noticeable
peak near \alphairac$\sim-2.5$ for the SINGS galaxies whereas the HCG
galaxies (and nuclei) are more evenly distributed from $\sim-4$ to
$\sim-1$. The larger number of HCG objects with \alphairac$>0$
compared to SINGs galaxies likely reflects the larger number of early
types in the former sample. However, morphological differences do not
account for all of the discrepancy, as a direct comparison of the
spirals in the two samples finds that while 11 of 47 (23\%) of the
SINGs spirals have \alphairac\ values in the range from --1.3 to 0,
none of the 17 HCG spirals do.  A Kolmogorov-Smirnov
analysis confirms the overall differences in the populations, giving only
0.12\% and 0.35\% probabilities that the SINGs galaxies are drawn from
the same distribution as the HCG nuclei and galaxies, respectively.
(The HCG nuclei and galaxies are consistent with eachother at 22.5\%
probability.) Hereafter, we will utilize the break in the HCG
\alphairac\ distribution to distinguish ``mid-infrared active''
(\alphairac$<0.0$) from ``mid-infrared quiescent''
(\alphairac$\ge0.0$) nuclei and galaxies.

We investigate this difference in more detail by plotting
\alphairac\ vs. \ltwofour\ for both the HCG and SINGS galaxies, coded
by galaxy morphology in Figure~\ref{fig:alphalum24}.  It is perhaps
most illuminating to compare the SINGS and HCG spiral galaxies
($\times$ symbols in Fig.~\ref{fig:alphalum24}).  While the two groups
cover a comparable range of \ltwofour, the HCG galaxies are
systematically redder in the mid-infrared, i.e., with more negative
values of \alphairac.  Within the HCG galaxies, morphology is
evidently a strong predictor of \alphairac, with elliptical and S0
galaxies almost exclusively populating the parameter space of
\alphairac$>0$.

We next consider the utility of measuring \frachd, the fraction of
\lfour\ attributable to non-stellar (i.e., hot dust) emission.  A plot
of \frachd\ vs. \leight\ shows a clear positive correlation in the HCG
population, with higher values of \leight\ corresponding to larger
values of \frachd, as shown in Figure~\ref{fig:fracd}a.  This
correlation is highly significant according to the non-parametric
Spearman's $\rho$ test which gives a probability, $P_{\rm S}$, of
$5.9\times10^{-7}$ for $\rho=0.67$ that these parameters are
uncorrelated.  While the HCG nuclei known to have optical
absorption-line spectra (filled circles) all have values of
\frachd$\sim0.0$ consistent with purely stellar emission and
\leight$\lesssim10^{28.5}$\lumin~Hz$^{-1}$, there is no clear
separation in this parameter space between galaxies identified as AGNs
or star-forming.  This is perhaps surprising, as one might expect AGNs
to have larger values of \frachd\ than star-forming galaxies because
of the typically hotter temperatures of dust emission associated with
harder ionizing continua.  This is true even for Seyfert~2 galaxies
that are typically assumed to have intrinsic absorption that blocks
the direct line of sight to the accretion disk
\citep[e.g.,][]{buchanan06}.  One source of hot dust emission in
star-forming galaxies that may contribute significantly at
$\sim4.5$\micron\ is from dust shells around AGB stars,
\citep[e.g.,][]{piovan}.

The five known low-ionization nuclear emission region (LINER) galaxies (open
squares), which can be powered by either star-formation or accretion
\citep[e.g.,][]{Eracleous02}, also do not lie in a distinct region
within this parameter space.  The recent mid-infrared spectroscopic
survey of \citet{sturm+06} found two distinct LINER populations
separable by infrared luminosity; the two populations can be divided
at $L_{\rm IR}\sim10^{10}$\lsun\ (where $L_{\rm IR}$ is integrated from
8--1000\micron).  In brief, infrared emission lines, continuum shape,
and PAH features indicate that infrared-luminous LINERs are more
closely linked to starburst galaxies, while infrared-faint LINERs were
identified primarily with AGNs.  However, the presence of high
ionization lines such as [\ion{O}{4}] seems to indicate that as many as
90\% of all LINERs in their sample contain AGNs, though accretion
power is not energetically dominant in the majority of cases.  The two
LINERs in our sample with no evidence for mid-infrared activity, 42a
and 62a, are known X-ray sources, and therefore likely do host AGNs.
Their low mid-infrared luminosities and weak emission-line
signatures in the optical indicate that the AGNs are not particularly
powerful.  Given that their values of
\mbox{$\nu$\leight} of 4.0$\times10^{8}$ and 6.6$\times10^{8}$\lsun\ 
are also the two lowest in the sample, they might be identified with
the infrared-faint LINERs of \citet{sturm+06}. The other three (16a,
16d, and 90d) may therefore be star-formation powered, though this
remains ambiguous without mid-infrared spectra.

Focusing on longer wavelengths, three galaxies show evidence for dust
emission at 24\micron\ that is not evident when only considering the
IRAC data.  In particular, a plot of \alphathree\ vs. \fracd\ reveals
a clear separation between galaxies at \fracd$\sim0.7$, see
Figure~\ref{fig:fracd}b.  With only one exception (62d), all known HCG
absorption-line galaxies have \fracd$<0.6$ and \alphathree$>0.5$.
None of the identified star-forming galaxies is in that region, while
two LINERs (42a and 62a) and one optical AGN (22a) are.  While all of the galaxies
identified as mid-infrared-active from their values of \alphairac$<0$
have \alphathree$<0.5$ and \fracd$>0.9$, an additional three
galaxies, the Seyfert~2s 16b and 61a and the absorption-line galaxy
62d, appear as 24\micron-excess sources.  An examination of their
mid-infrared SEDs (Fig.~\ref{fig:sed}) indicates that these galaxies
likely include a strong contribution from stellar photospheric emission
throughout most of the IRAC bands, and only at 24\micron, near a minimum
of the elliptical template SED, is there sufficient contrast to see
the low-luminosity dust continuum. Hereafter, we designate all sources
with \fracd$>0.7$ and \alphathree$<0.5$ as ``24\micron-active''.
 
Combining diagnostics of mid-infrared and 24\micron\ activity,
\alphathree\ vs. \alphairac\ is shown in Figure~\ref{fig:alphas},
coded by both \HI\ richness (as listed in column~9 of
Table~\ref{tab:groups}) and galaxy morphology. In this plot, it is
clear again that all nuclei designated as mid-infrared-active are also
24\micron-active, though the converse is not true for three objects.  The
dashed line represents \alphairac=\alphathree, and the elliptical
galaxy nuclei (with the exception of 62d) are all tightly clustered near
this line.  These objects are also typically in \HI-poor groups.
Nuclei from each of the most \HI-rich (Type~I) groups, 2, 16, 31,
and 61, have the most negative (reddest) \alphairac\ values.  This
implies a connection between the availability of \HI\ and 24\micron\
activity from both AGNs and star formation.  However, not all nuclei
within an \HI-rich group are 24\micron-active, though the fraction is
quite high ranging from 67 to 100\% for the four \HI-rich groups (see
Table~\ref{tab:groups}).  Evaluating the relationship between the
fraction of 24\micron-active (\fracd$>0.7$) nuclei and group \HI\
mass as shown in Figure~\ref{fig:gas}a, the non-parametric Spearman's
$\rho$ test gives a probability, $P_{\rm S}$, of only
$9.6\times10^{-4}$ for $\rho=0.83$ that these parameters are
uncorrelated.  Considering only mid-infrared-active galaxies, the
correlation is less significant ($\rho=0.72$; $P_{\rm
S}=8.0\times10^{-3}$), though still present.

Thirty-one of the 46 HCG galaxies have $H_2$ masses (including 14
upper limits) calculated from CO surveys \citep{verdes98}.  (Of these,
90bd and 31ace were unresolved in the radio images.)  In a plot of
$H_2$ vs. \ltwofour, the most mid-infrared luminous nuclei have
large amounts of $H_2$, however, there is quite a large scatter as
shown in Figure~\ref{fig:gas}b.  In fact, not all mid-infrared
luminous galaxies have large $H_2$ reservoirs.  However, more complete
and deeper molecular gas surveys are needed to explore this
relationship in further detail.

\subsection{Comparing Mid-Infrared Diagnostics for Integrated Galaxy
vs. Nuclear Light}

Given the proximity of our targets, the {\em Spitzer} angular
resolution offers a handle on investigating the difference between
nuclear and integrated galaxy parameters for identifying potential
nuclear activity.  This is particularly relevant for survey work,
where more distant objects are typically unresolved with {\em
Spitzer}.  Using the criterion that values of \alphairac$<0.0$
indicate mid-infrared activity, the galaxy and nuclear data for the
HCG sample are completely consistent (see Figure~\ref{fig:hists}b).
At longer wavelengths, however, there are some discrepancies.  In
particular,
\fracd$>0.7$ as the discriminant for a 24\micron\ excess yields some
differences depending on whether the integrated galaxy light or
nuclear light is considered.   Though all galaxies with
\fracd$(\rm nuc)>0.7$ also have \fracd$(\rm gal)>0.7$, four additional
galaxies --- 7b, 42a, 42b, and 90b --- have 24\micron\ excesses in
their integrated galaxy light.  This is also evident from visual
inspection of their SEDs as plotted in Figure~\ref{fig:sed}.  The values
of \fracd\ of the integrated galaxy vs. nuclear light are shown in
Figure~\ref{fig:galnuc}.  The four discrepant galaxies are all early type
(SB0, E3, SB0, and E0, respectively); the 24\micron\ excess is thus
evidence of low-level star formation occurring outside of the nuclei
in these galaxies.  If such objects were at a much greater distance
such as typical of large area surveys, the corresponding lack of spatial
resolution could therefore lead to inaccuracies in the nuclear
classification based on mid-infrared data alone.  No activity would be
missed, but some objects would be classified as 24-\micron\ active
(and therefore potentially hosting LLAGNs) where extranuclear 
star formation is in fact responsible.


\section{Summary and Conclusions}

Of the complete set of \total\ HCG galaxies presented in the {\em
Spitzer} imaging survey of J07, we have identified 25 with red
(\alphairac$<0.0$) mid-infrared continua.  All eight known,
spectroscopically identified star-forming galaxies (7a, 16c, 19c, 22c,
31ace, and 31b) are within this group. An additional three galaxies
(16b, 61a, and 62b --- all optically identified AGNs) are apparently
quiescent in the IRAC bands but show evidence for a 24\micron-excess
above a stellar continuum, bringing the total number of
24\micron-active HCG galaxies to 61\% (as all of the mid-infrared
active galaxies are also 24\micron-active).  This is higher than the
fraction of emission-line galaxies (including AGNs and star-forming
galaxies) identified in optical spectroscopic surveys of HCGs
\citep{coziol98a,coziol98b,shimada00}, and highlights the utility of
multiwavelength studies to capture all activity.  

\begin{figure*}
\plotone{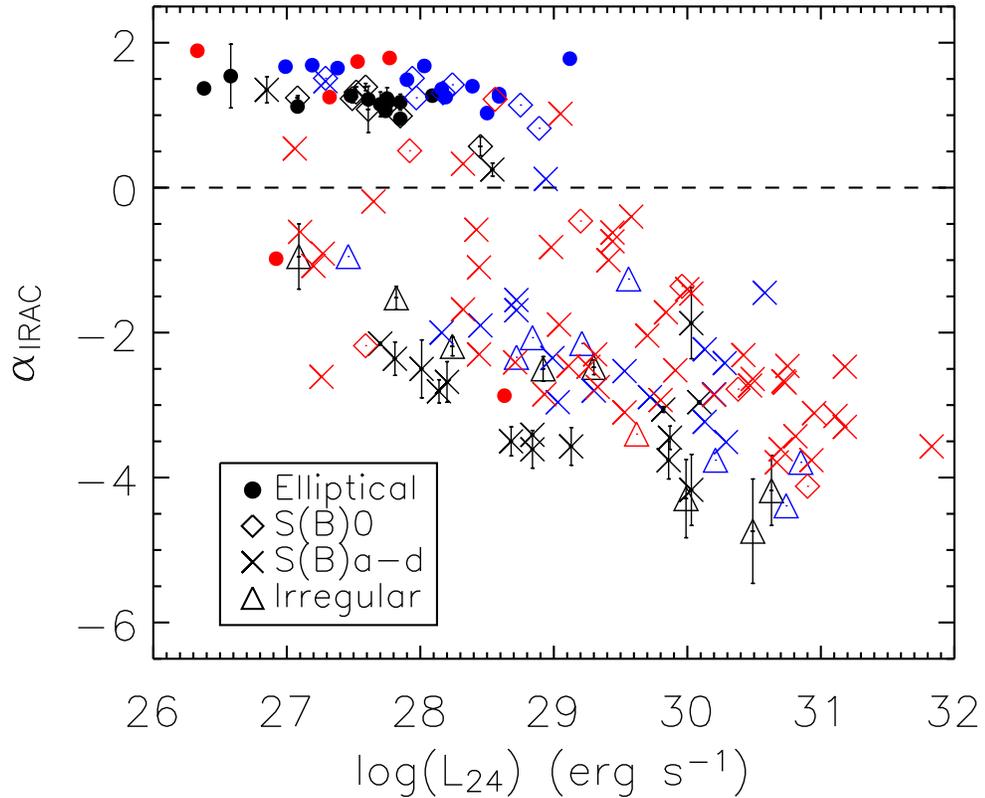}
\caption{A plot of \alphairac\ vs. \ltwofour\ for the HCG nuclei
(black symbols with error bars); HCG galaxies (blue symbols without
error bars) and SINGS comparison galaxy sample (red symbols without
error bars).  Datapoints have been coded by rough morphology classes
as indicated in the legend.  For a given \ltwofour, the HCG nuclei
tend to have redder (more negative values of \alphairac) mid-infrared
spectral slopes than the SINGS comparison sample.
\label{fig:alphalum24}
}
\end{figure*}

\begin{figure*}
\plottwo{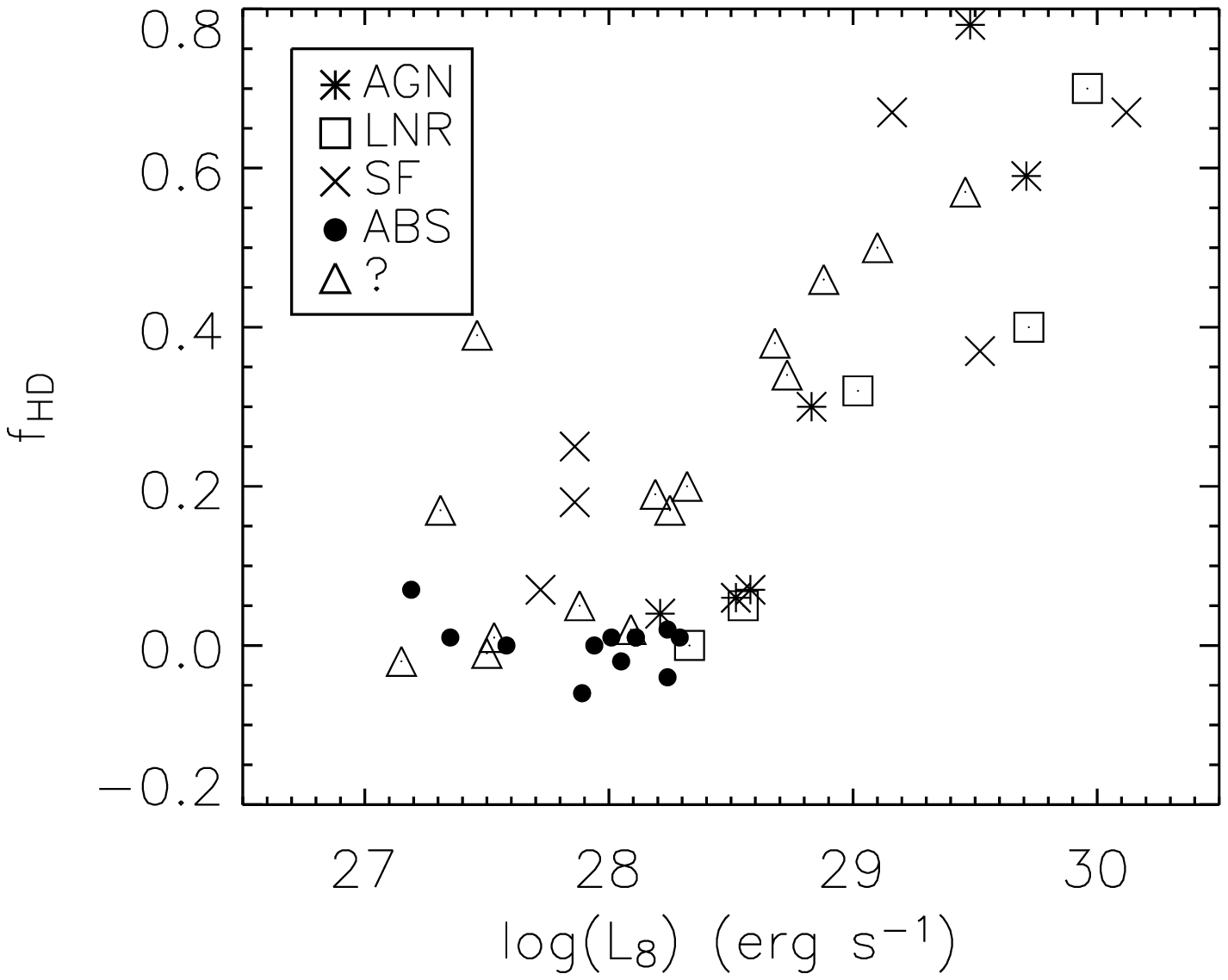}{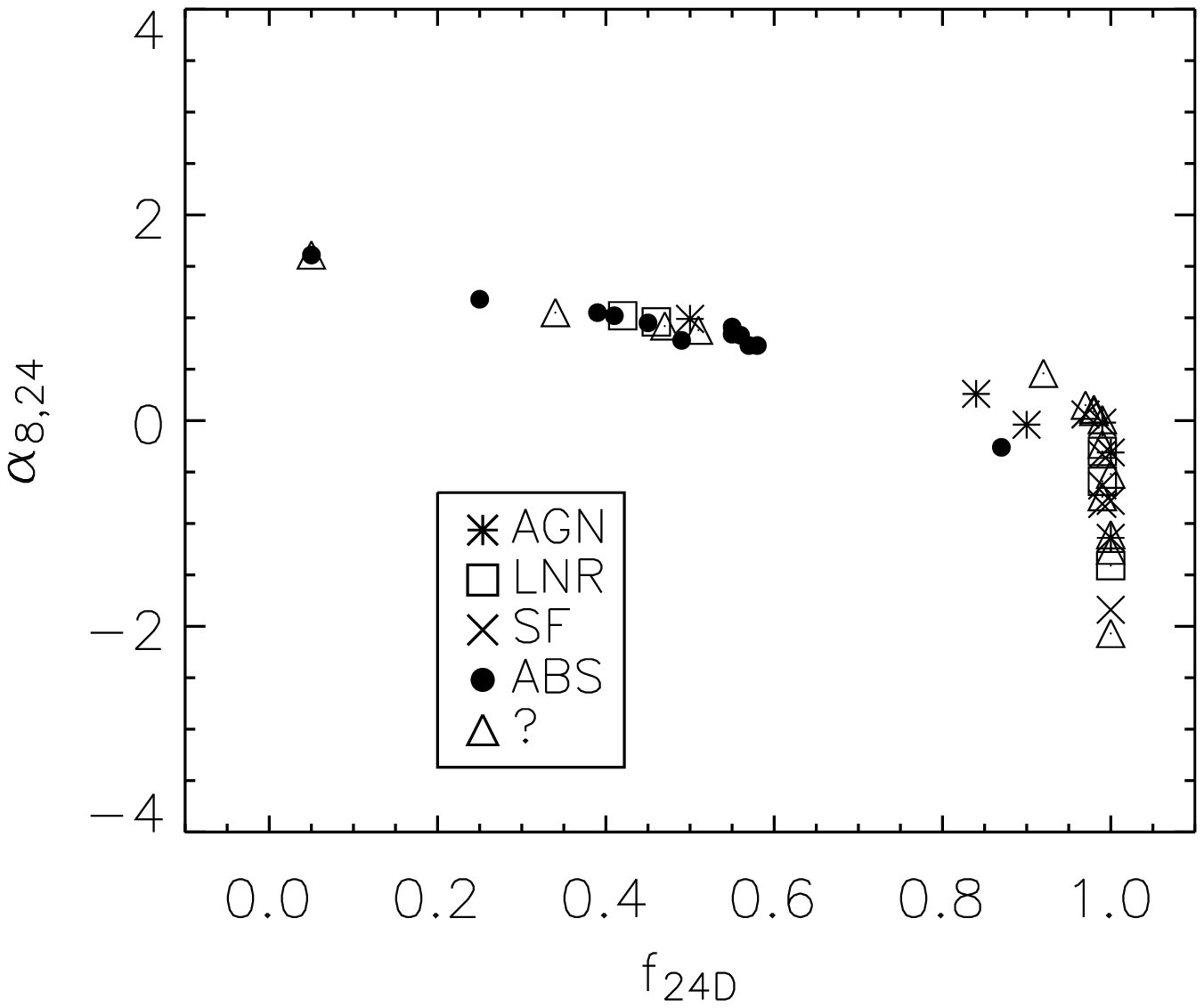}
\caption{({\em Left}) The fraction of hot dust emission at 4.5\micron,
  \frachd, vs. the 8\micron\ luminosity for the HCG nuclei coded by
  optical spectroscopic classification as listed in
  Table~\ref{tab:props} as AGNs (including Sy2 and AGN; asterisks),
  LINERS (open boxes), star-forming (including HII and ELG; $\times$'s),
  absorption-line (filled circles), and unclassified (open triangles).
  The optically classified absorption-line galaxies all have values of
  \frachd$\sim0$. ({\em Right}) The spectral slope between \leight\
  and \ltwofour, \alphathree, vs. the fraction of dust emission of
  24\micron, \fracd.  Symbols are coded as for the left panel. A
  24\micron\ excess can reveal low-level nuclear activity that is
  diluted at shorter wavelengths by stellar emission.
\label{fig:fracd}
}
\end{figure*}

\begin{figure*}
\plottwo{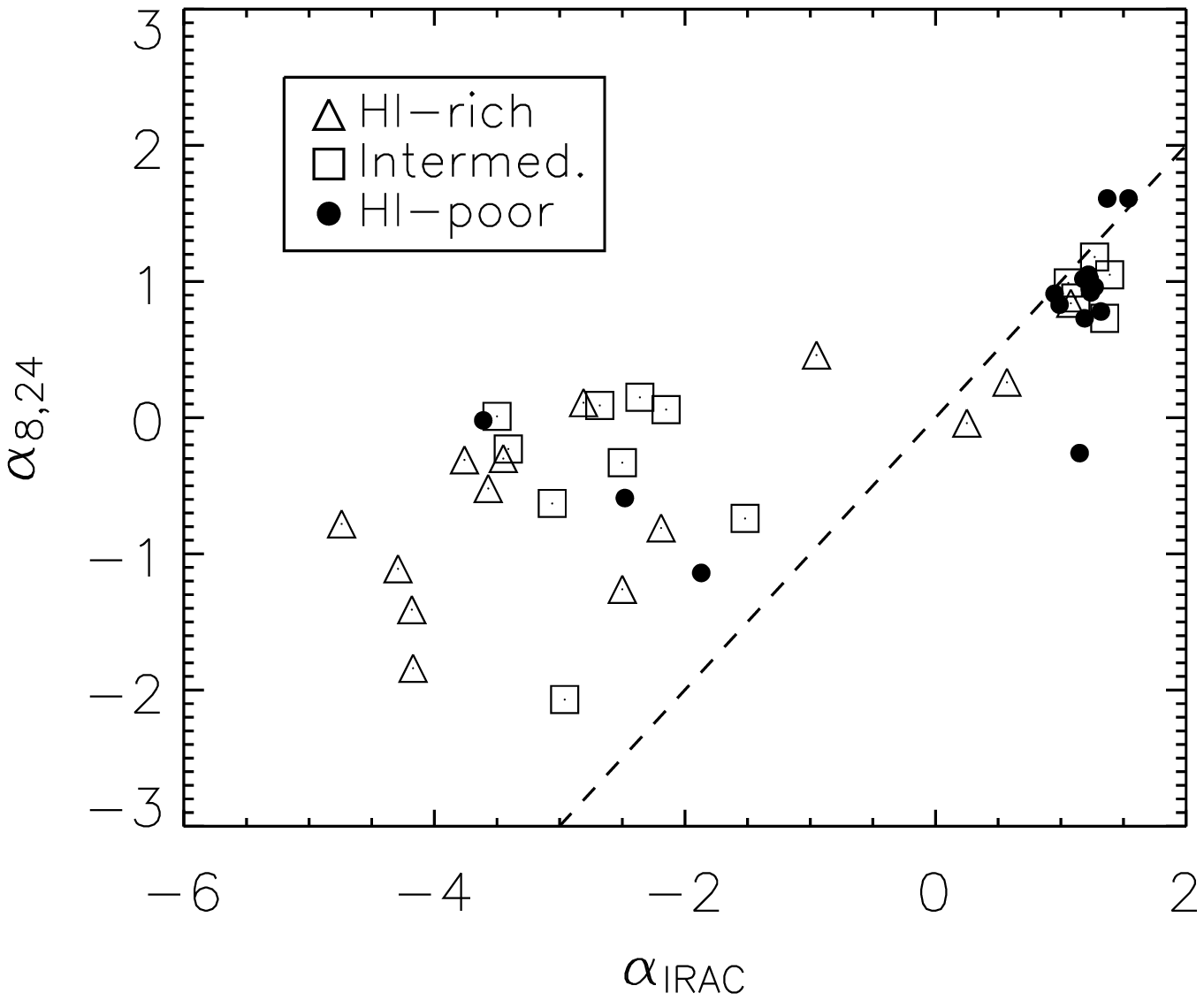}{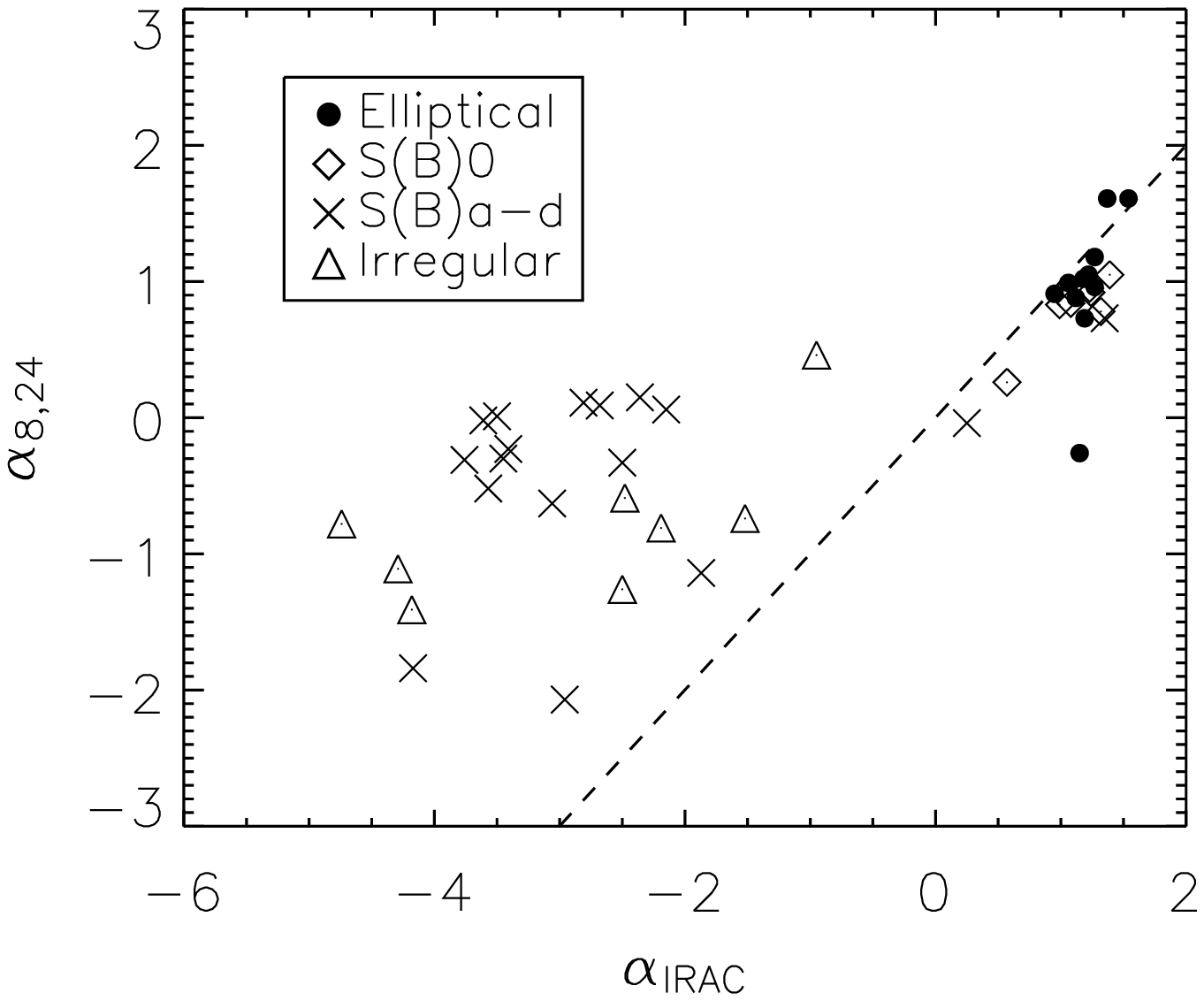}
\caption{({\em Left}) The 8-to-24\micron\ spectral index, \alphathree,
  vs. \alphairac. In this parameter space, nuclei with no evidence for
  mid-infrared activity are tightly clustered with values of
  \alphairac$>0.0$ and \alphathree$>0.5$; the dashed line marks
  \alphairac=\alphathree. The HCG nuclei have been coded according to
  their group type in terms of neutral hydrogen content: \HI-rich (Type~I; open triangles), 
  intermediate (Type~II; open boxes), and \HI-poor (Type~III; filled circles) as
  listed in Table~\ref{tab:groups}.
  ({\em Right}) The same plot as shown in the left-hand panel, with HCG
  nuclei coded according to morphological type as in 
  Figure~\ref{fig:alphalum24}.
\label{fig:alphas}
}
\end{figure*}

\begin{figure*}
\plottwo{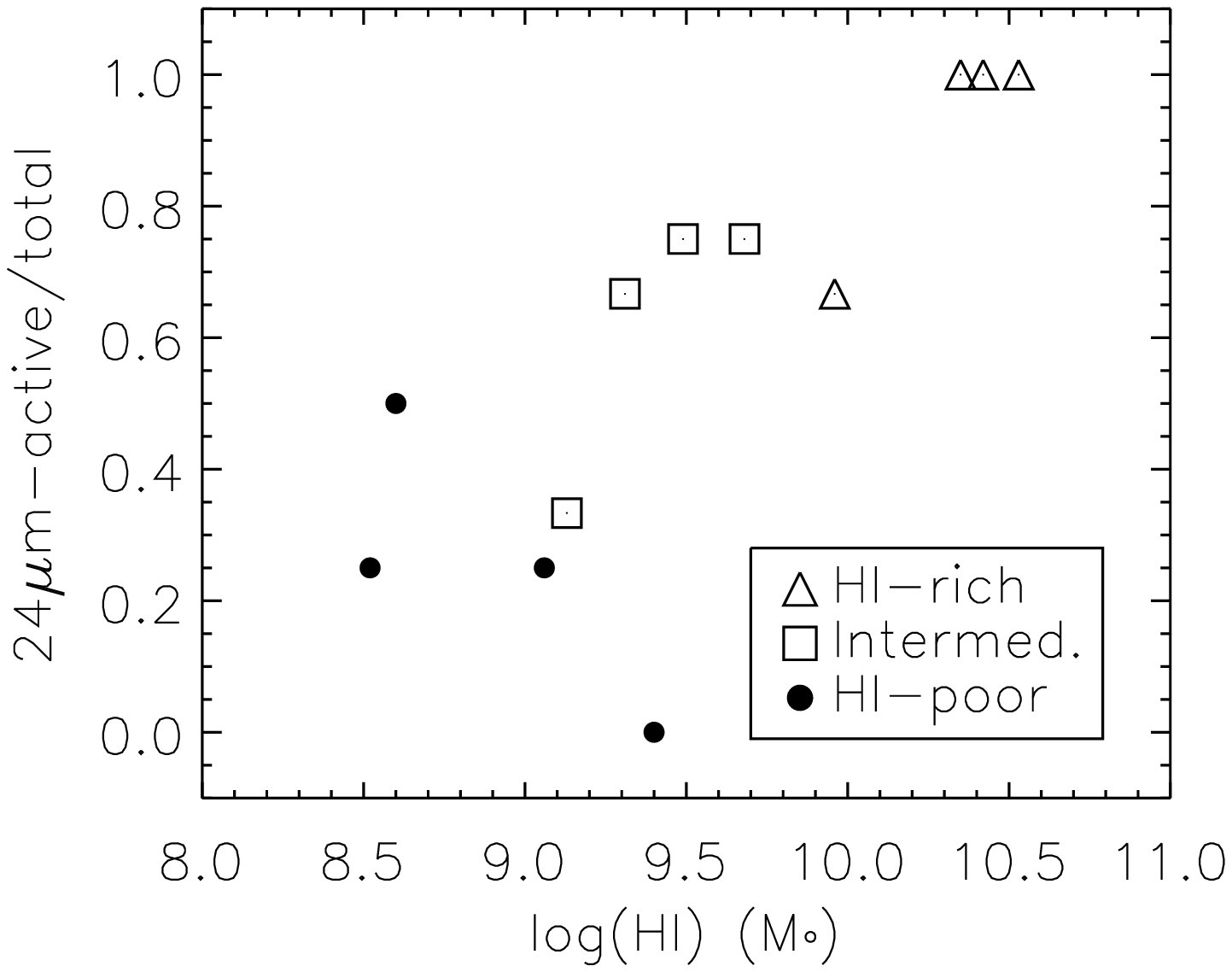}{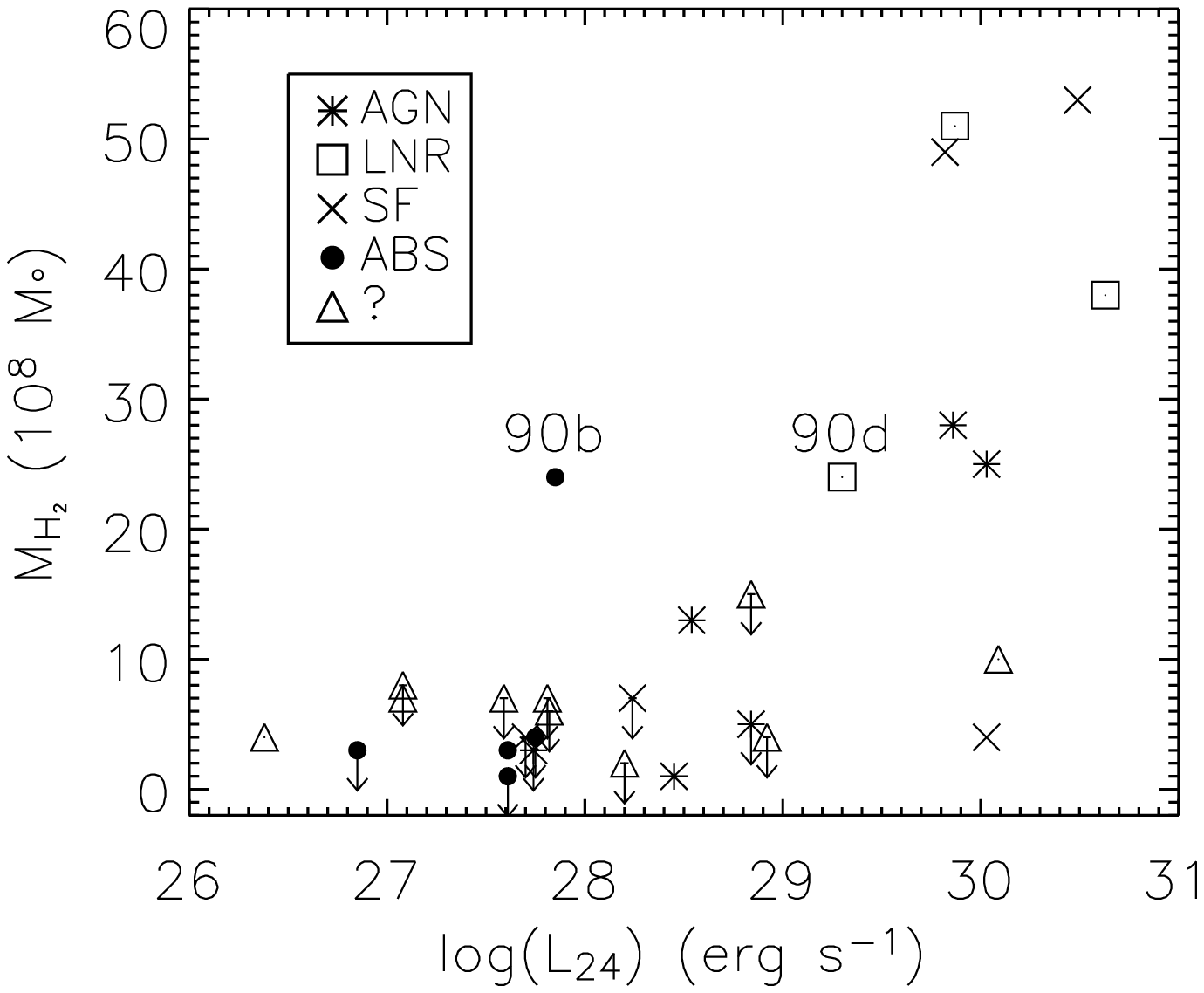}
\caption{({\em Left}) The fraction of 24\micron-active nuclei in a group vs. the
group mass in \HI.  Groups have been coded by \HI-richness as in
Figure~\ref{fig:alphas}a.
({\em Right})
The mass in molecular hydrogen, $M_{\rm H_2}$,
vs. \ltwofour\ for the 31 HCG galaxies with CO data from
\citet{verdes98}.  The galaxies 90b and 90d are labeled; they were
unresolved and so the combined $M_{\rm H_2}$ value is plotted for
both.  Galaxies with the most massive molecular hydrogen reserves
have nuclei among the most infrared luminous.
\label{fig:gas}
}
\end{figure*}

\begin{figure}
\plotone{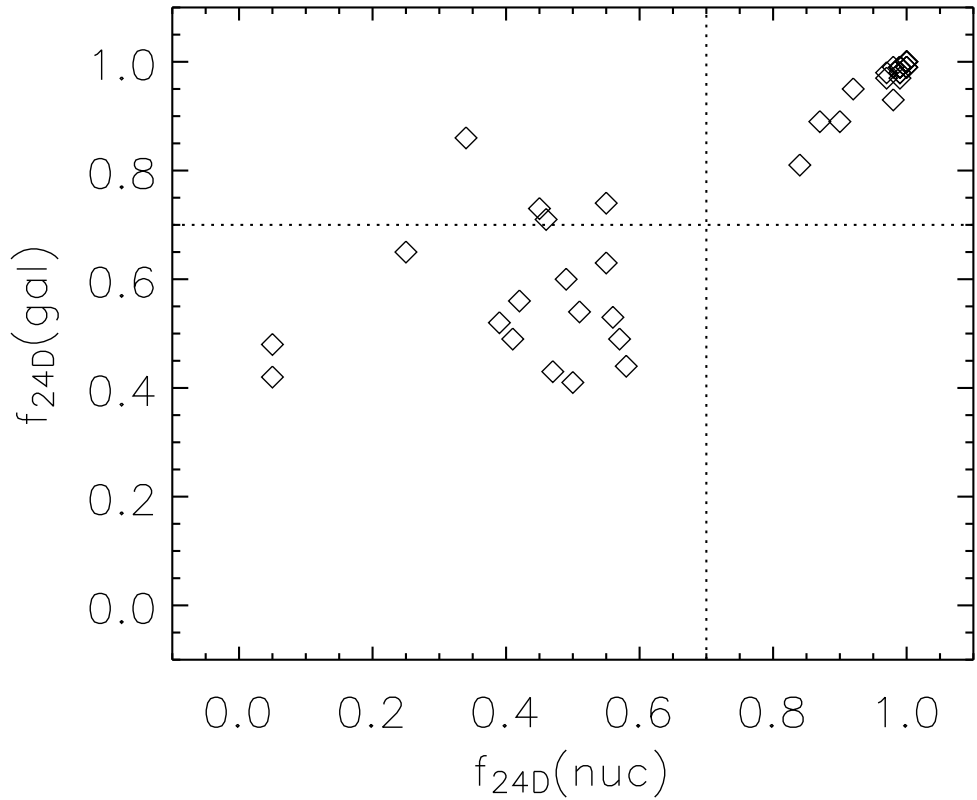}
\caption{The value of \fracd\ for the integrated HCG galaxy light
vs. the nuclear light.  The dotted lines indicate values of 0.7; all
objects with \fracd$>0.7$ are considered to have 24\micron\ excesses.
The four galaxies in the upper left section (7b, 42a, 42b, and 90b)
show evidence for 24\micron\ excesses in their integrated galaxy
emission but not in their nuclei indicating extranuclear
star-formation.
\label{fig:galnuc}
}
\end{figure}

Though the mid-infrared analysis presented here has the advantage of
being unambiguous in terms of identifying mid-infrared activity -- the
distribution of HCG nuclei is clearly bimodal in terms of both
\alphairac\ and \fracd\ -- distinguishing between accretion and star
formation as the dominant source of mid-infrared power remains
challenging.  The evident trend in Figure~\ref{fig:alphalum24}
between \alphairac\ and \ltwofour\ seen in mid-infrared active galaxy
nuclei could be interpreted as an increasing contribution from star
formation at higher \ltwofour\ and redder (more negative) values of
\alphairac.
Establishing this would be extremely powerful for interpreting
photometric data of fainter objects in wide field surveys.
However, the significant correlation between \frachd\ and \leight,
which might similarly be read to imply a stronger AGN component (because of the
greater hot dust contribution at 4.5\micron) at larger values of
\leight, is not generally consistent with the known nuclear identifications from
optical spectroscopy.  This implies that simple measures of the
steepness of the mid-infrared continuum are not diagnostic --
underscoring the difficulty of using mid-infrared photometry alone to
categorize objects.  Even with high signal-to-noise mid-infrared
spectroscopy, the underlying source of mid-infrared power can remain
ambiguous in some objects
\citep[e.g.,][]{weedman05}. Furthermore, the strong stellar
contamination at 3.6\micron\ in all of these galaxies means that
color-color diagnostics that use the ratio of 3.6 to 4.5\micron\
fluxes to isolate AGNs \citep[e.g.,][]{stern05} likely miss a large
fraction of them, particularly at lower infrared luminosities.
Selection criteria with a wider dynamic range in color will be less
sensitive to this effect \citep[e.g.,][]{lacy04}.  However, as noted
by J07, none of our sample would fulfill the AGN mid-infrared
color-color selection criteria of \citet{stern05} or
\citet{lacy04}, which were designed to target objects where the AGN
dominates the mid-infrared SED.  Our entire sample would also fail the
power-law AGN selection criterion of
\citet{donley07} that required a monotonic flux increase from 3.6 to
8.0\micron.

A few optically identified LLAGNs -- 22a, 42a, and 62a -- show no
evidence for any excess mid-infrared (including 24\micron) emission,
indicating that warm to hot
dust emission does not always accompany accretion onto a black hole.
However, without a reservoir of cold material, an optically detectable
AGN may be short-lived.  In any case, such a system is unlikely to
contribute significantly to the luminous energy budget of its host
galaxy.  Given their weak emission lines (which would be strongly
diluted by host galaxy light at larger distances; \altcite{moran02}),
these three galaxies may be identified with the optically dull but
X-ray detected AGNs found in clusters
\citep{martini06}.  The evolved nature of their host groups and the
X-ray detections of 42a and 62a in particular are consistent with
identifying these systems as mini-clusters.  In contrast, the less
evolved compact groups may be less likely to host X-ray bright,
optically quiet galaxies, as found in the loose group survey of
\citet{shen+07}.  At present, the lack of a complete suite of optical
spectroscopy and high quality X-ray observations of our galaxy sample
complicates definitively fitting the compact groups in context with
the AGN surveys in the loose group and cluster environments done to date.

The striking difference between the distribution of \alphairac\ values
for the HCG and SINGS galaxies supports the role of environment in
affecting HCG galaxies --- a conclusion consistent with the marked
difference in the mid-infrared color-color distribution of HCG and FLS
galaxies noted by J07. In particular, the strongly bimodal nature of
HCG nuclear (and galaxy) \alphairac\ values as well as the correlation
between group \HI\ content and 24\micron\ activity suggests that star
formation in a compact group galaxy, likely induced by interactions,
occurs in a burst that then exhausts its reservoir of cold gas, as
suggested previously from far-infrared
\citep{verdes98} and optical spectroscopic \citep{coziol98b} studies.  
The end result is a group dominated by mid-infrared quiescent
galaxies.  If local compact galaxy groups are analogous to the
building blocks of clusters in the early Universe, this implies that
the exhaustion of cold gas through enhanced star-formation occurs as a
result of interactions (though not necessarily mergers) prior to
cluster infall.  Any AGN activity in those galaxies at that point
would proceed in a low-luminosity, X-ray bright mode \citep{shen07}.
However, Tully-Fisher \citep{Mendes2003} and fundamental plane studies
\citep{delaRosa01} of HCG galaxies reveal that they are consistent
with control galaxy populations.  Therefore, there is no strong
evidence for a significant effect of the compact group environment on
the current states of galaxy morphology in such a significant way that
would alter the end states of galaxy evolution.

From a stellar population study of HCG ellipticals, \citet{delaRosa07}
concluded that these galaxies showed evidence for truncated star
formation in comparison to a (very small) sample of field ellipticals.
While this may be the case, the mechanism for star formation
truncation is unlikely to be extensive feedback from AGN
activity. None of the optically known AGNs in any HCG has sufficient
luminosity to clear a galaxy of its interstellar medium.  The sole
possible exception to this observation is 62a, where X-ray cavities in
the intragroup medium suggest that a relativistic jet has ejected a
sigificant amount of kinetic energy from a radiatively quiet AGN
\citep[e.g.,][]{morita+06,gu+07}.  However, this ongoing process likely
postdates the epoch of star formation in this evolved group.

\acknowledgements

We thank Justin Howell for helpful suggestions, and Karen Peterson for
contributions to an early version of this work.  We thank Paul Martini
and Varoujan Gorjian for their useful feedback.  This work is based on
observations made with the \spitzer\ Space Telescope, which is
operated by JPL/Caltech under a contract with NASA. Support for this
work was provided by NASA through an award issued by JPL/Caltech.
K.~E.~J. gratefully acknowledges partial support for this research for
this research provided by NSF/CAREER grant AST-0548103.  This project
has made use of the NASA/IPAC Extragalactic Database (NED) and the
Infrared Science Archive (IRSA), operated by JPL/Caltech, under
contract with NASA. This publication also makes use of 2MASS
(http://www.ipac.caltech.edu/2mass) data products.

{\it Facilities}\facility{\spitzer}\facility{2MASS}


\clearpage


\begin{thebibliography}{39}
\expandafter\ifx\csname natexlab\endcsname\relax\def\natexlab#1{#1}\fi
\expandafter\ifx\csname url\endcsname\relax
  \def\url#1{{\tt #1}}\fi
\expandafter\ifx\csname urlprefix\endcsname\relax\def\urlprefix{URL }\fi
\providecommand{\eprint}[2][]{\url{#2}}

\bibitem[\protect\astroncite{{Baron} \& {White}}{1987}]{baron}
{Baron}, E. \& {White}, S.~D.~M. 1987, \apj, 322, 585

\bibitem[Best et al.(2007)]{best07} 
Best, P.~N., von der Linden, A., Kauffmann, G., Heckman, T.~M., \&
Kaiser, C.~R.\ 2007, \mnras, 379, 894

\bibitem[\protect\astroncite{{Buchanan} et~al.}{2006}]{buchanan06}
{Buchanan}, C.~L., {Gallimore}, J.~F., {O'Dea}, C.~P., {Baum}, S.~A.,
{Axon}, D.~J., {Robinson}, A., {Elitzur}, M., \& {Elvis}, M. 2006,
\aj, 132, 401

\bibitem[\protect\astroncite{{Condon} et~al.}{1998}]{nvss_ref}
{Condon}, J.~J., {Cotton}, W.~D., {Greisen}, E.~W., {Yin}, Q.~F., {Perley},
  R.~A., {Taylor}, G.~B., \& {Broderick}, J.~J. 1998, \aj, 115, 1693

\bibitem[\protect\astroncite{{Coziol} et~al.}{1998{\natexlab{a}}}]{coziol98b}
{Coziol}, R., {de Carvalho}, R.~R., {Capelato}, H.~V., \& {Ribeiro}, A.~L.~B.
  1998{\natexlab{a}}, \apj, 506, 545

\bibitem[\protect\astroncite{{Coziol} et~al.}{1998{\natexlab{b}}}]{coziol98a}
{Coziol}, R., {Ribeiro}, A.~L.~B., {de Carvalho}, R.~R., \& {Capelato}, H.~V.
  1998{\natexlab{b}}, \apj, 493, 563

\bibitem[\protect\astroncite{{Dale} et~al.}{2007}]{dale+07}
{Dale}, D.~A., et~al. 2007, \apj, 655, 863

\bibitem[\protect\astroncite{{de Carvalho} et~al.}{1997}]{deCar97}
{de Carvalho}, R.~R., {Ribeiro}, A.~L.~B., {Capelato}, H.~V., \& {Zepf}, S.~E.
  1997, \apjs, 110, 1

\bibitem[\protect\astroncite{{de la Rosa} et~al.}{2007}]{delaRosa07}
{de la Rosa}, I.~G., {de Carvalho}, R.~R., {Vazdekis}, A., \& {Barbuy}, B.
  2007, \aj, 133, 330

\bibitem[\protect\astroncite{{de la Rosa} et~al.}{2001}]{delaRosa01}
{de la Rosa}, I.~G., {de Carvalho}, R.~R., \& {Zepf}, S.~E. 2001, \aj, 122, 93

\bibitem[\protect\astroncite{{Donley} et~al.}{2007}]{donley07}
{Donley}, J.~L., {Rieke}, G.~H., {Perez-Gonzalez}, P.~G., {Rigby}, J.~R., \&
  {Alonso-Herrero}, A. 2007, \apj, 660, 167

\bibitem[\protect\astroncite{{Dressler} et~al.}{1985}]{dressler85}
{Dressler}, A., Thompson, I.~B., \& Shectman, S.~A. 1985, \apj, 288, 481

\bibitem[\protect\astroncite{{Eracleous} et~al.}{2002}]{Eracleous02}
{Eracleous}, M., {Shields}, J.~C., {Chartas}, G., \& {Moran}, E.~C. 2002, \apj,
  565, 108

\bibitem[\protect\astroncite{{Gu} et~al.}{2007}]{gu+07}
{Gu}, J., {Xu}, H., {Gu}, L., {An}, T., {Wang}, Y., {Zhang}, Z., \& {Wu}, X.-P.
  2007, \apj, 659, 275

\bibitem[\protect\astroncite{{Hickson} et~al.}{1989}]{hickson89b}
{Hickson}, P., {Kindl}, E., \& {Auman}, J.~R. 1989, \apjs, 70, 687

\bibitem[\protect\astroncite{{Ho} et~al.}{1997{\natexlab{a}}}]{HoEtal97c}
{Ho}, L.~C., {Filippenko}, A.~V., \& {Sargent}, W.~L.~W. 1997{\natexlab{a}},
  \apjs, 112, 315

\bibitem[\protect\astroncite{{Ho} et~al.}{1997{\natexlab{b}}}]{HoEtal97a}
--- 1997{\natexlab{b}}, \apj, 487, 568

\bibitem[\protect\astroncite{Johnson et~al.}{2007}]{kelsey07}
Johnson, K.~E., Hibbard, J.~E., Gallagher, S.~C., Hornschemeier, A.~E., \&
  Charlton, J.~C. 2007, \aj, 134, 1522

\bibitem[\protect\astroncite{{Lacy} et~al.}{2004}]{lacy04}
{Lacy}, M., et~al. 2004, \apjs, 154, 166

\bibitem[\protect\astroncite{{Malhotra} et~al.}{2005}]{malhotra05}
{Malhotra}, S., et~al. 2005, \apj, 626, 666

\bibitem[\protect\astroncite{{Martini} et~al.}{2006}]{martini06}
{Martini}, P., {Kelson}, D.~D., {Kim}, E., {Mulchaey}, J.~S., \& Athey, A.~A.
  2006, \apj, 644, 116

\bibitem[\protect\astroncite{{Martini} et~al.}{2007}]{martini07}
{Martini}, P., {Mulchaey}, J.~S., \& {Kelson}, D.~D. 2007, \apj, 664, 761

\bibitem[\protect\astroncite{{Mendes de Oliveira} et~al.}{2003}]{Mendes2003}
{Mendes de Oliveira}, C., {Amram}, P., {Plana}, H., \& {Balkowski}, C. 2003,
  \aj, 126, 2635

\bibitem[\protect\astroncite{Moran et~al.}{2002}]{moran02}
Moran, E.~C., Filippenko, A.~V., \& Chornock, R. 2002, \apj, 579, L71

\bibitem[\protect\astroncite{{Morita} et~al.}{2006}]{morita+06}
{Morita}, U., {Ishisaki}, Y., {Yamasaki}, N.~Y., {Ota}, N., {Kawano}, N.,
  {Fukazawa}, Y., \& {Ohashi}, T. 2006, \pasj, 58, 719

\bibitem[Piovan et al.(2003)]{piovan} Piovan, L.,
Tantalo, R., \& Chiosi, C.\ 2003, \aap, 408, 559 

\bibitem[\protect\astroncite{{Richards} et~al.}{2006}]{ric+06}
{Richards}, G.~T., et~al. 2006, \apjs, 166, 470

\bibitem[\protect\astroncite{{Rubin} et~al.}{1991}]{rubin+91}
{Rubin}, V.~C., {Hunter}, D.~A., \& {Ford}, W.~K.~J. 1991, \apjs, 76, 153

\bibitem[\protect\astroncite{{Rudick} et~al.}{2006}]{rudick+06}
{Rudick}, C.~S., {Mihos}, J.~C., \& {McBride}, C. 2006, \apj, 648, 936

\bibitem[\protect\astroncite{Shen et~al.}{2007}]{shen07}
Shen, Y., Mulchaey, J.~S., Raychaudhury, S., Rasmussen, J., \& Ponman, T.~J.
  2007, \apj, 654, L115

\bibitem[\protect\astroncite{{Shen} et~al.}{2007}]{shen+07}
{Shen}, Y., {Mulchaey}, J.~S., {Raychaudhury}, S., {Rasmussen}, J., \&
  {Ponman}, T.~J. 2007, \apjl, 654, L115

\bibitem[\protect\astroncite{{Shimada} et~al.}{2000}]{shimada00}
{Shimada}, M., {Ohyama}, Y., {Nishiura}, S., {Murayama}, T., \& {Taniguchi}, Y.
  2000, \aj, 119, 2664

\bibitem[\protect\astroncite{{Silva} et~al.}{1998}]{grasil_ref}
{Silva}, L., {Granato}, G.~L., {Bressan}, A., \& {Danese}, L. 1998, \apj, 509,
  103

\bibitem[\protect\astroncite{{Spergel} et~al.}{2007}]{sperg07}
{Spergel}, D.~N., et~al. 2007, \apjs, 170, 377

\bibitem[\protect\astroncite{{Stern} et~al.}{2005}]{stern05}
{Stern}, D., et~al. 2005, \apj, 631, 163

\bibitem[\protect\astroncite{{Sturm} et~al.}{2006}]{sturm+06}
{Sturm}, E., et~al. 2006, \apjl, 653, L13

\bibitem[\protect\astroncite{{Verdes-Montenegro} et~al.}{1998}]{verdes98}
{Verdes-Montenegro}, L., {Yun}, M.~S., {Perea}, J., {del Olmo}, A., \& {Ho},
  P.~T.~P. 1998, \apj, 497, 89

\bibitem[\protect\astroncite{{Verdes-Montenegro} et~al.}{2001}]{verdes}
{Verdes-Montenegro}, L., {Yun}, M.~S., {Williams}, B.~A., {Huchtmeier}, W.~K.,
  {Del Olmo}, A., \& {Perea}, J. 2001, \aap, 377, 812

\bibitem[\protect\astroncite{{Weedman} et~al.}{2005}]{weedman05}
{Weedman}, D.~W., et~al. 2005, \apj, 633, 706

\bibitem[\protect\astroncite{{White} et~al.}{2003}]{white}
{White}, P.~M., {Bothun}, G., {Guerrero}, M.~A., {West}, M.~J., \& {Barkhouse},
  W.~A. 2003, \apj, 585, 739

\bibitem[\protect\astroncite{{White} et~al.}{1997}]{first_ref}
{White}, R.~L., {Becker}, R.~H., {Helfand}, D.~J., \& {Gregg}, M.~D. 1997,
  \apj, 475, 479

\end{thebibliography}
\begin{deluxetable}{lcccccrccr}
\tablecolumns{10}
\tablewidth{0pt}
\tablecaption{Hickson Compact Group Properties
\label{tab:groups}
}
\tablehead{
\colhead{HCG} &
\colhead{$\bar{v}$\tablenotemark{a}} &
\multicolumn{4}{c}{Gal. Morph\tablenotemark{b}} &
\colhead{$\log(M_{\rm HI})$\tablenotemark{c}} &
\colhead{Evol.} &
\colhead{\HI} &
\colhead{} \\
\colhead{Name} &
\colhead{(\kms)} &
\colhead{E} &
\colhead{S(B)0} &
\colhead{S(B)a--d} &
\colhead{Irr} &
\colhead{(\msun)} &
\colhead{Stage\tablenotemark{d}} &
\colhead{Type\tablenotemark{e}} &
\colhead{$f_{\rm IR,act}$\tablenotemark{f}} 
}
\startdata
2    &  4309  & 0 & 0 & 2 & 1  &  10.53 &  early       & I   & 3/3   \\
7    &  4233  & 0 & 1 & 3 & 0  &   9.68 &  early	      &	II   & 3/4   \\
16   &  3957  & 0 & 0 & 2 & 2  &  10.42 &  int	      &	I  & 3(1)/4   \\
19   &  4245  & 1 & 0 & 2 & 0  &   9.31 &  early/int   &	II   & 2/3   \\
22   &  2686  & 1 & 0 & 2 & 0  &   9.13 &  early	      &	II   & 1/3   \\
31   &  4094  & 0 & 0 & 4 & 2  &  10.35 &  int	      &	I  & 6/6   \\
42   &  3976  & 3 & 1 & 0 & 0  &   9.40 &  late	      &	III   & 0/4   \\
48\tablenotemark{g}   &  3162  & 2 & 1 & 1 & 0  &   8.52 &  late	      &	III   & 1/4   \\
59   &  4058  & 1 & 0 & 1 & 1  &   9.49 &  early/int   &	II   & 3/4   \\
61   &  3907  & 0 & 2 & 1 & 0  &   9.96 &  early/int   &	I   & 1(1)/3   \\
62   &  4122  & 2 & 2 & 0 & 0  &   9.06 &  late	      &	III   & 0(1)/4   \\
90   &  2644  & 2 & 0 & 1 & 1  &   8.60 &  int         & III   & 2/4   \\
\enddata
\tablenotetext{a}{Average recession velocity for all of the accordant compact
group galaxies.}
\tablenotetext{b}{Taken from \citet{hickson89b}; more detailed
morphologies are listed in Table~\ref{tab:props}.}
\tablenotetext{c}{Mass ($\log(M)$ in \msun) of neutral hydrogen \citep{verdes}.}
\tablenotetext{d}{A qualitative determination of evolutionary stage
based on group galaxy morphologies and X-ray detection of an
intragroup medium based on \citet{verdes}.}
\tablenotetext{e}{\HI\  type as measured and defined by
\citet{kelsey07}. Key: (I) = \HI\ rich ($\log(M_{\rm HI})/\log(M_{\rm
dyn})>0.9$); (II) = intermediate ($\log(M_{\rm HI})/\log(M_{\rm
dyn})=0.8$--0.9); (III) \HI\ poor ($\log(M_{\rm HI})/\log(M_{\rm
dyn})<0.8$).}
\tablenotetext{f}{Fraction of ``mid-infared active'' (\alphairac$<0$)
nuclei in each group.  Numbers in parentheses refer to
``24\micron-active'' nuclei with \fracd$>0.7$ and \alphairac$>0$.}
\tablenotetext{g}{Based on the velocities of its four galaxies (2267,
2437, 4381, 4361 \kms), HCG 48 may be more appropriately characterized
as two pairs rather than a proper compact group.}
\end{deluxetable}
\clearpage
\begin{deluxetable}{lrrrrrrrr}
\tablecolumns{9}
\tablewidth{0pt}
\tablecaption{Nuclear Fluxes for Accordant HCG Galaxies\tablenotemark{a}
\label{tab:fluxes}
}
\tablehead{
\colhead{HCG} &
\multicolumn{3}{c}{2MASS} &
\multicolumn{4}{c}{IRAC} &
\colhead{MIPS} \\
\colhead{Name} &
\colhead{$J$} &
\colhead{$H$} &
\colhead{$K$} &
\colhead{3.6} &
\colhead{4.5} &
\colhead{5.7} &
\colhead{8.0} &
\colhead{24}  \\
}
\startdata
2a    & $ 4.6\pm0.5$ & $5.6 \pm0.6$ & $4.3 \pm0.4$ & $3.2 \pm0.3$ & $2.2 \pm0.2$ & $6.0 \pm0.6	$ & $16.5\pm1.7$ & $29.3\pm2.9$ \\
2b    & $ 9.4\pm0.9$ & $10.8\pm1.1$ & $9.3 \pm0.9$ & $7.4 \pm0.7$ & $5.4 \pm0.5$ & $20.2\pm2.0	$ & $61.3\pm6.1$ & $212 \pm21.2$ \\
2c    & $ 2.2\pm0.2$ & $2.6 \pm0.3$ & $2.1 \pm0.2$ & $1.2 \pm0.1$ & $0.7 \pm0.1$ & $1.5	\pm0.2	$ & $3.4 \pm0.3$ & $3.0 \pm0.3$ \\
7a    & $30.7\pm3.1$ & $40.2\pm4.0$ & $36.4\pm3.6$ & $20.8\pm2.1$ & $13.4\pm1.3$ & $27.7 \pm2.8	$ & $74.5\pm7.5$ & $150 \pm15.0$ \\
7b    & $22.1\pm2.2$ & $26.7\pm2.7$ & $21.7\pm2.2$ & $10.8\pm1.1$ & $6.3 \pm0.6$ & $4.6	\pm0.5	$ & $2.9 \pm0.3$ & $0.9 \pm0.1$ \\
7c    & $ 4.6\pm0.5$ & $5.4 \pm0.5$ & $4.9 \pm0.5$ & $2.8 \pm0.3$ & $1.8 \pm0.2$ & $4.1	\pm 0.4	$ & $12.2\pm1.2$ & $15.7\pm1.6$ \\
7d    & $ 2.9\pm0.3$ & $3.3 \pm0.3$ & $2.6 \pm0.3$ & $1.5 \pm0.2$ & $0.9 \pm0.1$ & $1.5	\pm0.2	$ & $4.0 \pm0.4$ & $3.6 \pm0.4$ \\
16a   & $43.5\pm4.4$ & $56.0\pm5.6$ & $49.6\pm5.0$ & $30.0\pm3.0$ & $19.5\pm2.0$ & $49.0 \pm4.9	$ & $136 \pm13.6$ & $191\pm19.1$ \\
16b   & $33.0\pm3.3$ & $41.1\pm4.1$ & $34.8\pm3.3$ & $16.6\pm1.7$ & $9.8 \pm1.0$ & $8.8	\pm0.9	$ & $8.5 \pm0.8	$ & $8.9\pm0.9$ \\
16c   & $28.4\pm2.8$ & $36.8\pm3.7$ & $35.6\pm3.6$ & $31.5\pm3.2$ & $22.6\pm2.3$ & $103	\pm10.3	$ & $332 \pm33.2 $ & $790\pm79$ \\
16d   & $23.9\pm2.4$ & $31.8\pm3.2$ & $31.5\pm3.2$ & $24.6\pm2.5$ & $21.2\pm2.1$ & $74.8\pm7.5	$ & $226 \pm26.0 $ & $1090\pm109$ \\
19a   & $19.4\pm1.9$ & $23.6\pm2.4$ & $19.1\pm1.9$ & $9.1 \pm0.9$ & $5.3 \pm0.5$ & $3.9\pm0.4	$ & $2.6 \pm0.3	$ & $0.7\pm0.7$ \\
19b   & $ 3.6\pm0.4$ & $4.4 \pm0.4$ & $3.9 \pm0.4$ & $2.3 \pm0.2$ & $1.5 \pm0.2$ & $3.9\pm0.4	$ & $10.8\pm1.1	$ & $10.7\pm1.1$ \\
19c   & $ 1.0\pm0.1$ & $1.5 \pm0.2$ & $1.2 \pm0.1$ & $0.6 \pm0.1$ & $0.4 \pm0.1$ & $0.6\pm0.1	$ & $1.6 \pm0.2	$ & $2.3 \pm0.2$ \\
22a   & $57.8\pm5.8$ & $71.8\pm7.2$ & $60.0\pm6.0$ & $30.1\pm3.0$ & $16.8\pm1.7$ & $13.5\pm1.4	$ & $9.3 \pm0.9	$ & $3.1 \pm0.3$ \\
22b   & $ 6.3\pm0.6$ & $7.9 \pm0.8$ & $6.1 \pm0.6$ & $3.1 \pm0.3$ & $1.9 \pm0.2$ & $1.5\pm0.2	$ & $0.9 \pm0.1	$ & $0.4 \pm0.4$ \\
22c   & $ 3.2\pm0.3$ & $3.7 \pm0.4$ & $2.9 \pm0.3$ & $1.4 \pm0.2$ & $0.9 \pm0.1$ & $1.5\pm0.2	$ & $3.0 \pm0.3	$ & $2.8 \pm0.3$ \\
31ace & $ 4.3\pm0.4$ & $4.7 \pm0.5$ & $4.3 \pm0.4$ & $3.7 \pm0.4$ & $3.1 \pm0.3$ & $11.0\pm1.1	$ & $32.9\pm3.3	$ & $256 \pm26$ \\
31b   & $ 1.7\pm0.2$ & $1.7 \pm0.2$ & $1.5 \pm0.2$ & $0.8 \pm0.1$ & $0.5 \pm0.1$ & $0.8\pm0.1	$ & $1.7 \pm0.2	$ & $4.2 \pm0.4$ \\
31g   & $ 3.9\pm0.4$ & $4.2 \pm0.4$ & $3.7 \pm0.4$ & $1.9 \pm0.2$ & $1.2 \pm0.1$ & $2.4\pm0.2	$ & $4.9 \pm0.5	$ & $20.0\pm2.0$ \\
31q   & $ 0.9\pm0.1$ & $1.1 \pm0.1$ & $0.9 \pm0.1$ & $0.4 \pm0.1$ & $0.3 \pm0.1$ & $0.3\pm0.1	$ & $0.5 \pm0.1	$ & $0.3 \pm0.3$ \\
42a   & $63.1\pm6.3$ & $79.1\pm8.0$ & $66.5\pm6.7$ & $33.0\pm3.3$ & $18.9\pm1.9$ & $14.4\pm1.4	$ & $9.3 \pm0.9	$ & $3.2 \pm0.3$ \\
42b   & $16.6\pm1.7$ & $19.9\pm2.0$ & $16.3\pm1.6$ & $7.8 \pm0.8$ & $4.6 \pm0.5$ & $3.3\pm0.3	$ & $2.3 \pm0.2	$ & $0.8 \pm0.1$ \\
42c   & $23.2\pm2.3$ & $28.7\pm2.9$ & $23.5\pm2.4$ & $11.1\pm1.1$ & $6.6 \pm0.7$ & $5.1\pm0.5	$ & $3.4 \pm0.3	$ & $1.5 \pm0.2$ \\
42d   & $ 5.0\pm0.5$ & $6.1 \pm0.6$ & $4.9 \pm0.5$ & $2.3 \pm0.2$ & $1.4 \pm0.1$ & $1.2\pm0.1	$ & $0.6 \pm0.1	$ & $0.1 \pm0.1$ \\
48a   & $49.9\pm5.0$ & $62.4\pm6.2$ & $53.4\pm5.2$ & $24.8\pm2.5$ & $14.3\pm1.4$ & $11.4\pm1.1	$ & $7.2 \pm0.7	$ & $2.3 \pm0.2$ \\
48b   & $ 9.9\pm1.0$ & $11.4\pm1.1$ & $9.4 \pm0.9$ & $5.6 \pm0.6$ & $3.6 \pm0.4$ & $9.9\pm1.0	$ & $27.5\pm2.8 $ & $28.0\pm2.8$ \\
48c   & $ 9.8\pm1.0$ & $12.2\pm1.2$ & $10.1\pm1.0$ & $4.7 \pm0.5$ & $2.8 \pm0.3$ & $2.1\pm0.2	$ & $1.4 \pm0.1	$ & $0.5 \pm0.1$ \\
48d   & $ 4.6\pm0.5$ & $5.9 \pm0.6$ & $4.7 \pm0.5$ & $2.2 \pm0.2$ & $1.3 \pm0.1$ & $0.9\pm0.1	$ & $0.6 \pm0.1	$ & $0.1 \pm0.1$ \\
59a   & $10.5\pm1.1$ & $12.5\pm1.3$ & $10.7\pm1.1$ & $6.3 \pm0.6$ & $5.5 \pm0.6$ & $11.4\pm1.1	$ & $29.0\pm2.9	$ & $292 \pm29.2$ \\
59b   & $ 5.5\pm0.6$ & $6.5 \pm0.7$ & $5.4 \pm0.5$ & $2.4 \pm0.2$ & $1.5 \pm0.2$ & $1.1\pm0.1	$ & $0.8 \pm0.1	$ & $0.3 \pm0.1$ \\
59c   & $ 1.8\pm0.2$ & $2.2 \pm0.2$ & $1.6 \pm0.4$ & $0.8 \pm0.1$ & $0.5 \pm0.1$ & $1.0\pm0.1	$ & $1.9 \pm0.2	$ & $1.6 \pm0.2$ \\
59d   & $ 0.7\pm0.1$ & $0.8 \pm0.1$ & $0.5 \pm0.1$ & $0.4 \pm0.1$ & $0.3 \pm0.1$ & $0.4\pm0.1	$ & $0.7 \pm0.1	$ & $1.6 \pm0.2$ \\
61a   & $43.9\pm4.4$ & $60.7\pm6.1$ & $47.2\pm4.7$ & $23.1\pm2.3$ & $14.0\pm1.4$ & $11.4\pm1.1	$ & $10.1\pm1.0	$ & $7.6 \pm0.8$ \\
61c   & $21.8\pm2.2$ & $36.5\pm3.7$ & $33.6\pm3.4$ & $22.9\pm2.3$ & $16.1\pm1.6$ & $45.8\pm4.6	$ & $134 \pm13.4 $ & $190\pm19.0$ \\
61d   & $17.4\pm1.7$ & $23.3\pm2.3$ & $17.5\pm1.8$ & $8.6 \pm0.9$ & $5.2 \pm0.5$ & $3.4\pm0.3	$ & $2.8 \pm0.3	$ & $1.1 \pm0.1$ \\
62a   & $36.3\pm3.6$ & $45.2\pm4.5$ & $38.0\pm3.8$ & $17.8\pm1.8$ & $10.2\pm1.0$ & $8.1\pm0.8	$ & $5.3 \pm0.5	$ & $1.7 \pm0.2$ \\
62b   & $28.0\pm2.8$ & $33.8\pm3.4$ & $28.0\pm2.8$ & $13.0\pm1.3$ & $7.5 \pm0.8$ & $5.9\pm0.6	$ & $4.3 \pm0.4	$ & $1.7 \pm0.2$ \\
62c   & $15.4\pm1.5$ & $18.1\pm1.8$ & $14.1\pm1.4$ & $6.5 \pm0.7$ & $4.0 \pm0.4$ & $2.8\pm0.3	$ & $1.9 \pm0.2	$ & $0.8 \pm0.1$ \\
62d   & $ 6.2\pm0.6$ & $7.3 \pm0.7$ & $6.0 \pm0.6$ & $2.8 \pm0.3$ & $1.7 \pm0.2$ & $1.4\pm0.1	$ & $0.9 \pm0.1	$ & $1.2 \pm0.1$ \\
90a   & $43.0\pm4.3$ & $65.1\pm6.5$ & $68.3\pm6.8$ & $58.8\pm5.9$ & $59.2\pm5.9$ & $120\pm12.0	$ & $173 \pm17.3$ &  $\cdots$ \\
90b   & $69.6\pm7.0$ & $87.3\pm8.7$ & $71.2\pm7.1$ & $34.1\pm3.4$ & $19.8\pm2.0$ & $15.7\pm1.6	$ & $11.6\pm1.2$ & $4.2 \pm0.4$ \\
90c   & $54.2\pm5.4 $ & $66.9\pm6.7$ & $54.2\pm5.4$ & $26.9\pm2.7$ & $15.3\pm1.5$ & $11.4\pm1.1	$ & $7.7 \pm0.8$ & $2.4 \pm0.2$ \\
90d   & $36.1\pm3.6$ & $48.7\pm4.9$ & $43.2\pm4.3$ & $24.0\pm2.4$ & $15.1\pm1.5$ & $26.2\pm2.6	$ & $60.3\pm6.0$ & $117 \pm11.7$ \\
\enddata
\tablenotetext{a}{All fluxes (in units of mJy) for a 7''-radius
aperture centered on the nucleus; no aperture corrections have been
applied.  Near-infrared bandpasses are labeled by filter name, and
mid-infrared bandpasses by wavelength.}
\end{deluxetable}
\clearpage
\begin{deluxetable}{lrrrrrrrrrrrl}
\tablecolumns{13}
\tablewidth{0pt}
\tablecaption{Properties of HCG Galaxy Nuclei
\label{tab:props}
}
\tablehead{
\colhead{HCG} &
\colhead{Gal.} &
\colhead{Nuc.} &
\colhead{$M_{\rm H_2}$\tablenotemark{c}} &
\colhead{} &
\colhead{} &
\multicolumn{4}{c}{$\log(L_{\nu})$ (\lumin~Hz$^{-1}$)} &
\colhead{} &
\colhead{} &
\colhead{} \\
\colhead{Name} &
\colhead{Morph.\tablenotemark{a}} &
\colhead{Class.\tablenotemark{b}} &
\colhead{($10^8$~M$_{\odot}$)} &
\colhead{\alphairac\tablenotemark{d}} &
\colhead{\alphathree\tablenotemark{e}} &
\colhead{$J$\tablenotemark{f}} &
\colhead{4.5\tablenotemark{f}} &
\colhead{8\tablenotemark{f}} &
\colhead{24\tablenotemark{f}}  &
\colhead{\frachd\tablenotemark{g}} &
\colhead{\fracd\tablenotemark{g}} &
\colhead{References\tablenotemark{h}} \\
}
\startdata
   2a & SBd  &        R  &  $\cdots$ & $ -3.57\pm 0.26$ & $ -0.52$ &  28.33 & 28.03 &   28.88 &  29.13  & $  0.46$ &1.00& 1  \\  
    2b & cI  &        R  &  $\cdots$ & $ -4.29\pm 0.54$ & $ -1.11$ &  28.64 & 28.43 &   29.46 &  29.99  & $  0.57$ &1.00& 1 \\
   2c & SBc  &        ?  &  $\cdots$ & $ -2.81\pm 0.16$ & $  0.11$ &  28.01 & 27.53 &   28.19 &  28.14  & $  0.19$ &0.98&  \\
    7a & Sb  &      HII  &  $49.2  $ & $ -3.06\pm 0.04$ & $ -0.63$ &  29.14 & 28.79 &   29.52 &  29.82  & $  0.37$ &0.99& 2 \\
   7b & SB0  &        ?  &  $<7.3  $ & $  1.39\pm 0.05$ & $  1.05$ &  28.99 & 28.44 &   28.09 &  27.59  & $  0.02$ &0.34&  \\
   7c & SBc  &        ?  &  $15.0  $ & $ -3.41\pm 0.01$ & $ -0.23$ &  28.31 & 27.92 &   28.73 &  28.84  & $  0.34$ &0.99&  \\
   7d & SBc  &        ?  &  $<2.5  $ & $ -2.68\pm 0.28$ & $  0.09$ &  28.11 & 27.61 &   28.25 &  28.20  & $  0.17$ &0.98&  \\
 16a & SBab  &  LNR,X,R  &  $51.3  $ & $ -3.45\pm 0.16$ & $ -0.30$ &  29.23 & 28.90 &   29.72 &  29.87  & $  0.40$ &0.99& 1, 3, 6\\
  16b & Sab  &      Sy2  &  $13.5  $ & $  0.25\pm 0.09$ & $ -0.04$ &  29.11 & 28.57 &   28.52 &  28.54  & $  0.06$ &0.90& 3 \\
   16c & Im  &     SBNG  &  $53.7  $ & $ -4.74\pm 0.72$ & $ -0.78$ &  29.04 & 28.98 &   30.12 &  30.49  & $  0.67$ &1.00& 3 \\
   16d & Im  &      LNR  &  $38.0  $ & $ -4.18\pm 0.48$ & $ -1.41$ &  28.97 & 28.94 &   29.96 &  30.63  & $  0.70$ &1.00& 3 \\
   19a & E2  &      ABS  &  $\cdots$ & $  1.27\pm 0.00$ & $  1.18$ &  28.94 & 28.36 &   28.05 &  27.48  & $ -0.02$ &0.25& 3\\
  19b & Scd  &        ?  &  $\cdots$ & $ -3.50\pm 0.20$ & $  0.01$ &  28.21 & 27.85 &   28.68 &  28.68  & $  0.38$ &0.99& 4 \\
  19c & Sdm  &      ELG  &  $\cdots$ & $ -2.50\pm 0.40$ & $ -0.33$ &  27.66 & 27.26 &   27.86 &  28.01  & $  0.25$ &0.99&  \\
   22a & E2  &     dSy2  &  $<3.0  $ & $  1.06\pm 0.08$ & $  0.99$ &  29.01 & 28.47 &   28.21 &  27.74  & $  0.04$ &0.50& 3 \\
   22b & Sa  &      ABS  &  $<3.6  $ & $  1.35\pm 0.18$ & $  0.73$ &  28.05 & 27.52 &   27.19 &  26.85  & $  0.07$ &0.57& 4 \\
 22c & SBcd  &      ELG  &  $<4.0  $ & $ -2.15\pm 0.02$ & $  0.06$ &  27.76 & 27.21 &   27.72 &  27.70  & $  0.07$ &0.97& 4 \\
31ace & Sdm  &      HII  &  $4.4   $ & $ -4.17\pm 0.49$ & $ -1.84$ &  28.25 & 28.14 &   29.16 &  30.03  & $  0.67$ &1.00& 2 \\
   31b & Sm  &      HII  &  $<7.2  $ & $ -2.19\pm 0.13$ & $ -0.81$ &  27.85 & 27.33 &   27.86 &  28.24  & $  0.18$ &0.99& 2 \\
   31g & Im  &        ?  &  $<4.1  $ & $ -2.50\pm 0.17$ & $ -1.26$ &  28.21 & 27.71 &   28.32 &  28.92  & $  0.20$ &1.00&  \\
   31q & Im  &        ?  &  $\cdots$ & $ -0.95\pm 0.45$ & $  0.46$ &  27.57 & 27.09 &   27.31 &  27.09  & $  0.17$ &0.92&  \\
   42a & E3  &   dLNR,X  &  $\cdots$ & $  1.27\pm 0.07$ & $  0.96$ &  29.39 & 28.86 &   28.55 &  28.09  & $  0.05$ &0.46& 3, 6 \\
  42b & SB0  &      ABS  &  $\cdots$ & $  1.23\pm 0.06$ & $  0.95$ &  28.81 & 28.24 &   27.94 &  27.49  & $ -0.00$ &0.45& 4 \\
   42c & E2  &      ABS  &  $\cdots$ & $  1.19\pm 0.06$ & $  0.73$ &  28.96 & 28.40 &   28.11 &  27.76  & $  0.01$ &0.58& 4 \\
   42d & E2  &      ABS  &  $\cdots$ & $  1.54\pm 0.44$ & $  1.61$ &  28.29 & 27.73 &   27.35 &  26.58  & $  0.01$ &0.05& 4 \\
   48a & E2  &      ABS  &  $<4.8  $ & $  1.23\pm 0.15$ & $  1.02$ &  29.09 & 28.54 &   28.24 &  27.75  & $  0.02$ &0.41& 4 \\
   48b & Sc  &        X  &  $<5.9  $ & $ -3.61\pm 0.26$ & $ -0.02$ &  28.39 & 27.97 &   28.83 &  28.84  & $  0.30$ &0.99&  \\
  48c & S0a  &        ?  &  $7.7   $ & $  1.24\pm 0.03$ & $  0.92$ &  28.38 & 27.83 &   27.53 &  27.08  & $  0.01$ &0.47&  \\
   48d & E1  &        ?  &  $4.1   $ & $  1.37\pm 0.07$ & $  1.61$ &  28.06 & 27.50 &   27.15 &  26.38  & $ -0.02$ &0.05&  \\
   59a & Sa  &        ?  &  $10.2  $ & $ -2.96\pm 0.02$ & $ -2.07$ &  28.63 & 28.37 &   29.10 &  30.09  & $  0.50$ &1.00&  \\
   59b & E0  &        ?  &  $<8.8  $ & $  1.12\pm 0.08$ & $  0.88$ &  28.35 & 27.78 &   27.50 &  27.08  & $ -0.01$ &0.51&  \\
   59c & Sc  &        ?  &  $<7.2  $ & $ -2.36\pm 0.23$ & $  0.15$ &  27.87 & 27.33 &   27.88 &  27.81  & $  0.05$ &0.97&  \\
   59d & Im  &        ?  &  $<6.6  $ & $ -1.52\pm 0.16$ & $ -0.74$ &  27.46 & 27.09 &   27.46 &  27.82  & $  0.39$ &0.99&  \\
  61a & S0a  &    Sy2,R  &  $1.7   $ & $  0.57\pm 0.13$ & $  0.26$ &  29.22 & 28.72 &   28.58 &  28.45  & $  0.07$ &0.84& 2, 5 \\
  61c & Sbc  &    AGN,R  &  $28.2  $ & $ -3.76\pm 0.26$ & $ -0.31$ &  28.92 & 28.81 &   29.71 &  29.86  & $  0.59$ &1.00& 2, 5 \\
   61d & S0  &      ABS  &  $<1.9  $ & $  1.08\pm 0.32$ & $  0.84$ &  28.82 & 28.28 &   28.01 &  27.61  & $  0.01$ &0.55& 2, 6 \\
   62a & E3  &   dLNR,X  &  $\cdots$ & $  1.18\pm 0.11$ & $  1.02$ &  29.19 & 28.62 &   28.33 &  27.85  & $ -0.00$ &0.42& 2, 3 \\
   62b & S0  &    ABS,X  &  $\cdots$ & $  0.99\pm 0.00$ & $  0.83$ &  29.07 & 28.49 &   28.24 &  27.85  & $ -0.04$ &0.56& 2, 4 \\
   62c & S0  &      ABS  &  $\cdots$ & $  1.32\pm 0.07$ & $  0.78$ &  28.81 & 28.21 &   27.89 &  27.52  & $ -0.06$ &0.49& 2, 4 \\
   62d & E2  &      ABS  &  $\cdots$ & $  1.15\pm 0.17$ & $ -0.26$ &  28.42 & 27.85 &   27.58 &  27.70  & $  0.00$ &0.87& 4 \\
   90a & Sa  &  Sy2,X,R  &  $25.7 $ & $ -1.87\pm 0.49$ & $ -1.14$ &
28.87 & 29.02 &   29.48 &  30.03  & $  0.78$ & 1.00 & 3 \\
   90b\tablenotemark{i} & E0  &    ABS,X  &  $24.8  $ & $  0.95\pm 0.00$ & $  0.91$ &  29.08 & 28.53 &   28.29 &  27.85  & $  0.01$ &0.55& 4 \\
   90c & E0  &      ABS  &  $3.0   $ & $  1.22\pm 0.01$ & $  1.05$ &  28.97 & 28.41 &   28.11 &  27.61  & $  0.01$ &0.39& 4 \\  
   90d\tablenotemark{i} & Im  &    LNR,R  &  $24.8  $ & $ -2.48\pm 0.10$ & $ -0.59$ &  28.80 & 28.42 &   29.02 &  29.30  & $  0.32$ &0.99& 3 \\

\enddata
\tablenotetext{a}{Galaxy morphologies are taken from the 2-color imaging survey of \citet{hickson89b}.}
\tablenotetext{b}{Published diagnostics of nuclear type, primarily from optical spectroscopy. 
Key: ABS = absorption-line galaxies; AGN = active galactic nucleus; 
(d)Sy2 = (dwarf) Seyfert~2; X = X-ray source; R = radio source; (d)LNR = (dwarf) 
low-ionization nuclear emission region; SBNG = starburst nucleated galaxy; HII = strong \ion{H}{2} emitter;
ELG = emission-line galaxy; ? = unknown classification.}
\tablenotetext{c}{Values of $M_{\rm H_2}$ are from Table~1 of the CO survey of \citet{verdes98}.}
\tablenotetext{d}{The spectral slope determined from a power-law model fit to the 4.5, 5.7, and 8\micron\ monochromatic luminosities ($L_{\nu}\propto\nu^{\alpha}$).}  
\tablenotetext{e}{The spectral slope measured between the 8 and 24\micron\ monochromatic luminosities.}
\tablenotetext{f}{The logarithm of the monochromatic luminosities (\lumin~Hz$^{-1}$) at rest-frame 1.2 ($J$), 4.5, 8, and 24\micron.}  
\tablenotetext{g}{Fraction of 4.5 and 24\micron\ emission from dust
calculated by subtracting the observed flux from that expected from an elliptical galaxy template normalized to the $J$ and $H$-band emission.}
\tablenotetext{h}{References for the nuclear classifications.  Key: (1) National Radio Astronomical Observatory Very Large Array Sky Survey \citep[NVSS;][]{nvss_ref}; (2) optical spectroscopy \citep{shimada00}; (3) optical spectroscopy \citep{coziol98a}; (4) optical spectroscopy \citep{deCar97}; (5) Faint Images of the Radio Sky at 20 cm \citep[FIRST;][]{first_ref}; (6) NASA/IPAC Extragalactic Database (NED; nedwww.ipac.caltech.edu).} 
\tablenotetext{i}{Because the 24\micron\ photometry is not availabel, the values of \ltwofour\ and \fracd\ for 90a have been
determined by extrapolating from the IRAC photometry.}
\tablenotetext{j}{The galaxies 90b and 90d were unresolved in the CO survey of \citet{verdes98}, and the quoted $M_{\rm H_2}$ values for each galaxy are the total flux for both.}
\end{deluxetable}
\clearpage
\thispagestyle{empty}
\setlength{\voffset}{-28mm}
\clearpage
\setlength{\voffset}{0mm}
{\plotone{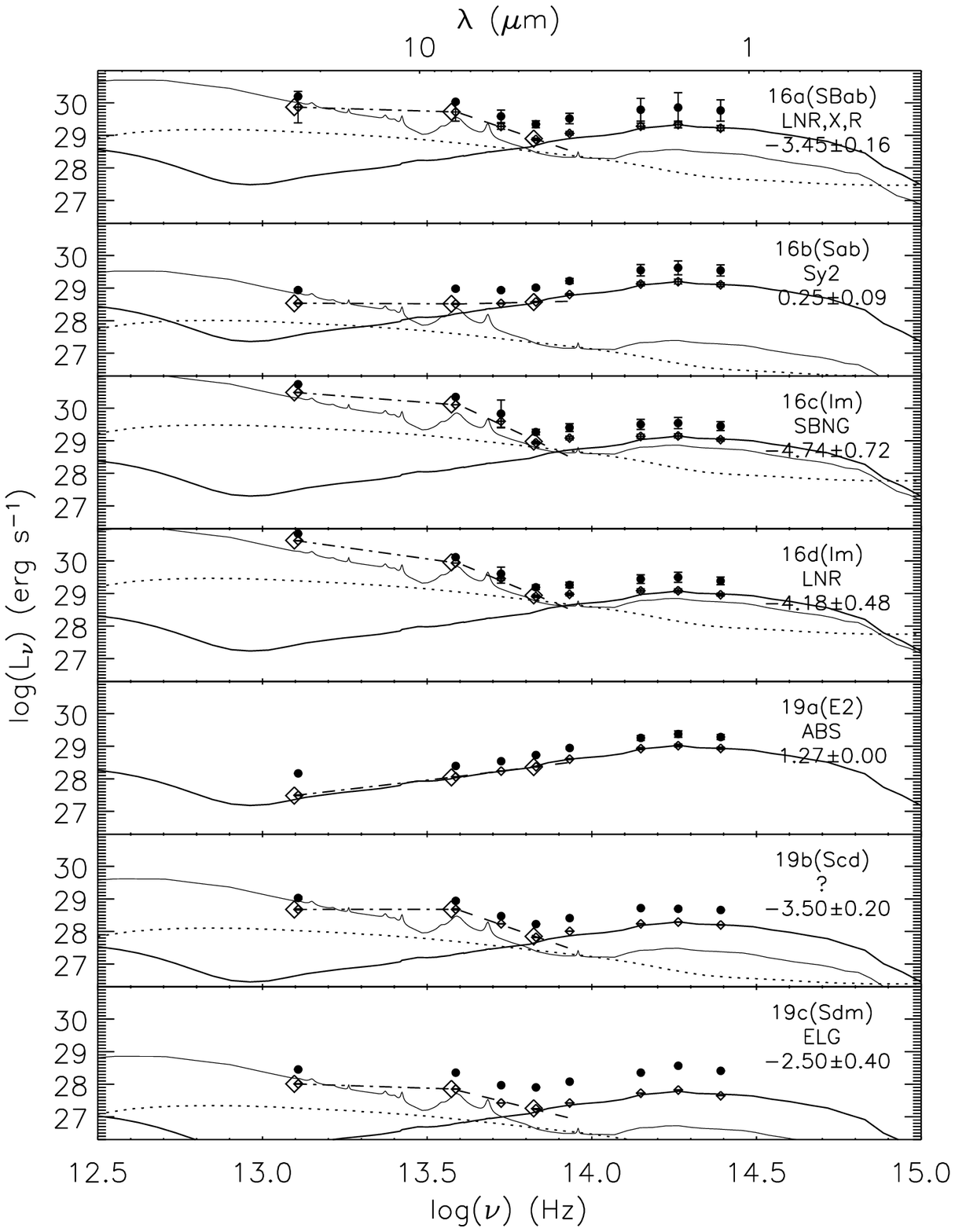}}\\
\centerline{Fig. 1. --- Continued.}
\clearpage
{\plotone{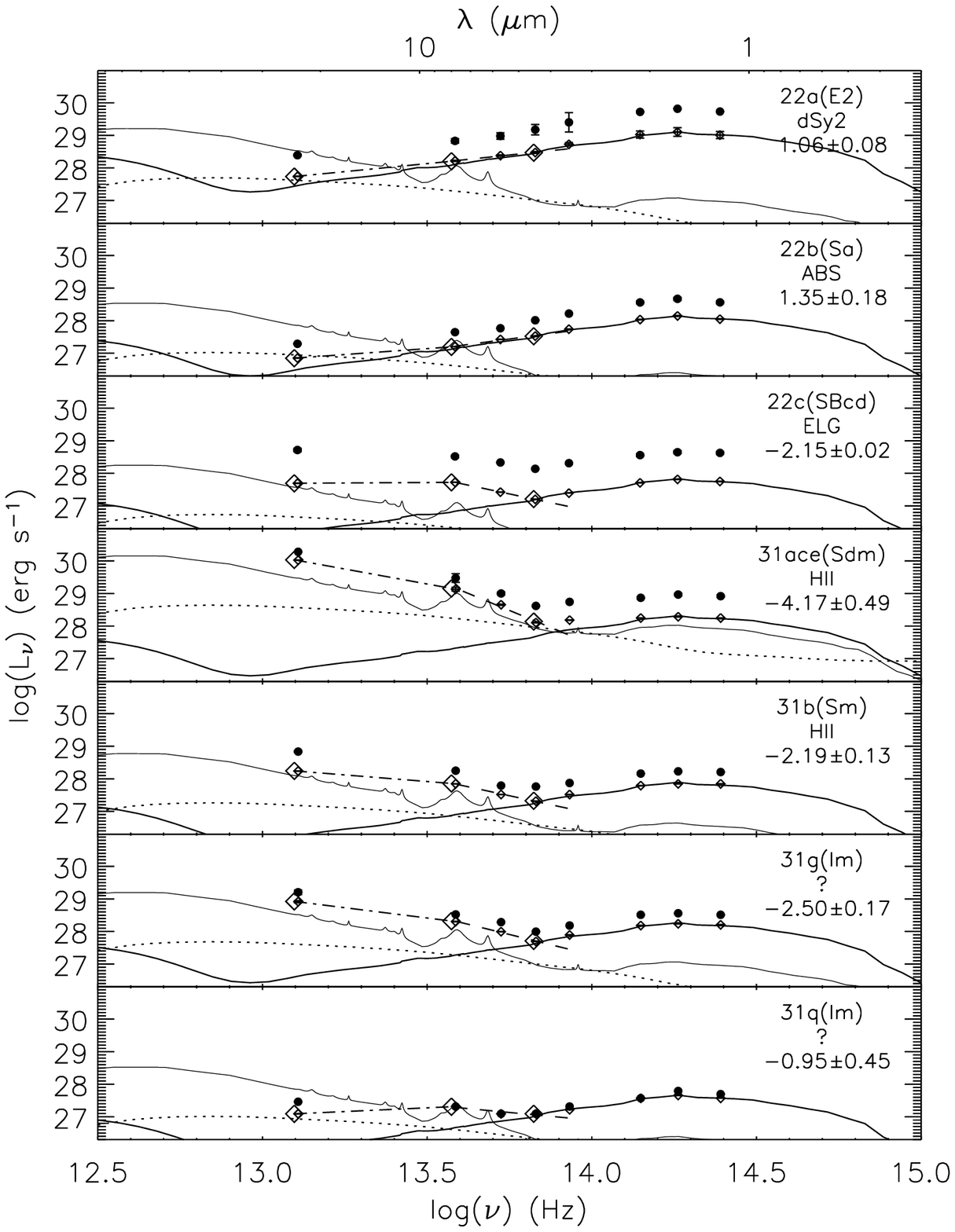}}\\
\centerline{Fig. 1. --- Continued.}
\clearpage
{\plotone{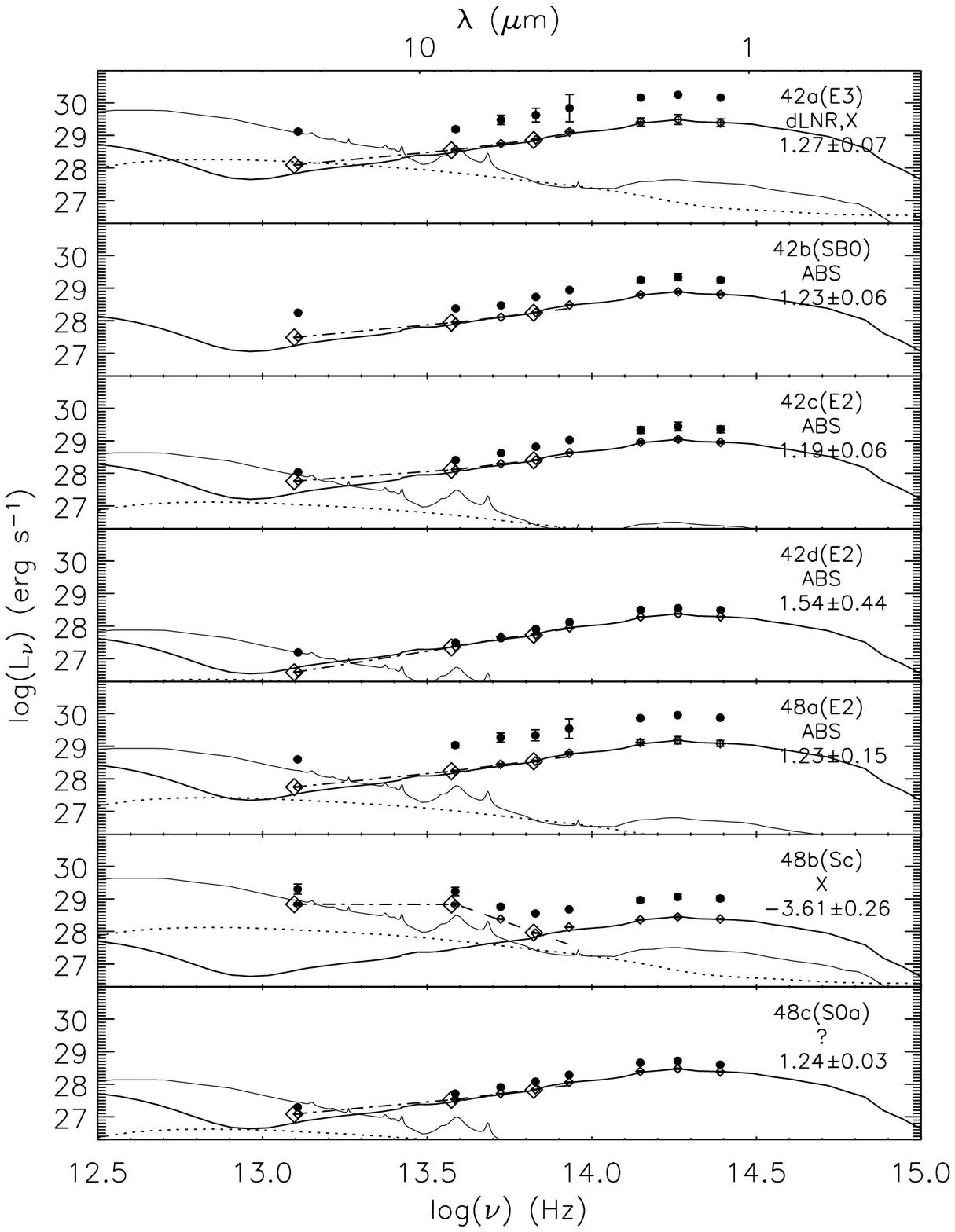}}\\
\centerline{Fig. 1. --- Continued.}
\clearpage
{\plotone{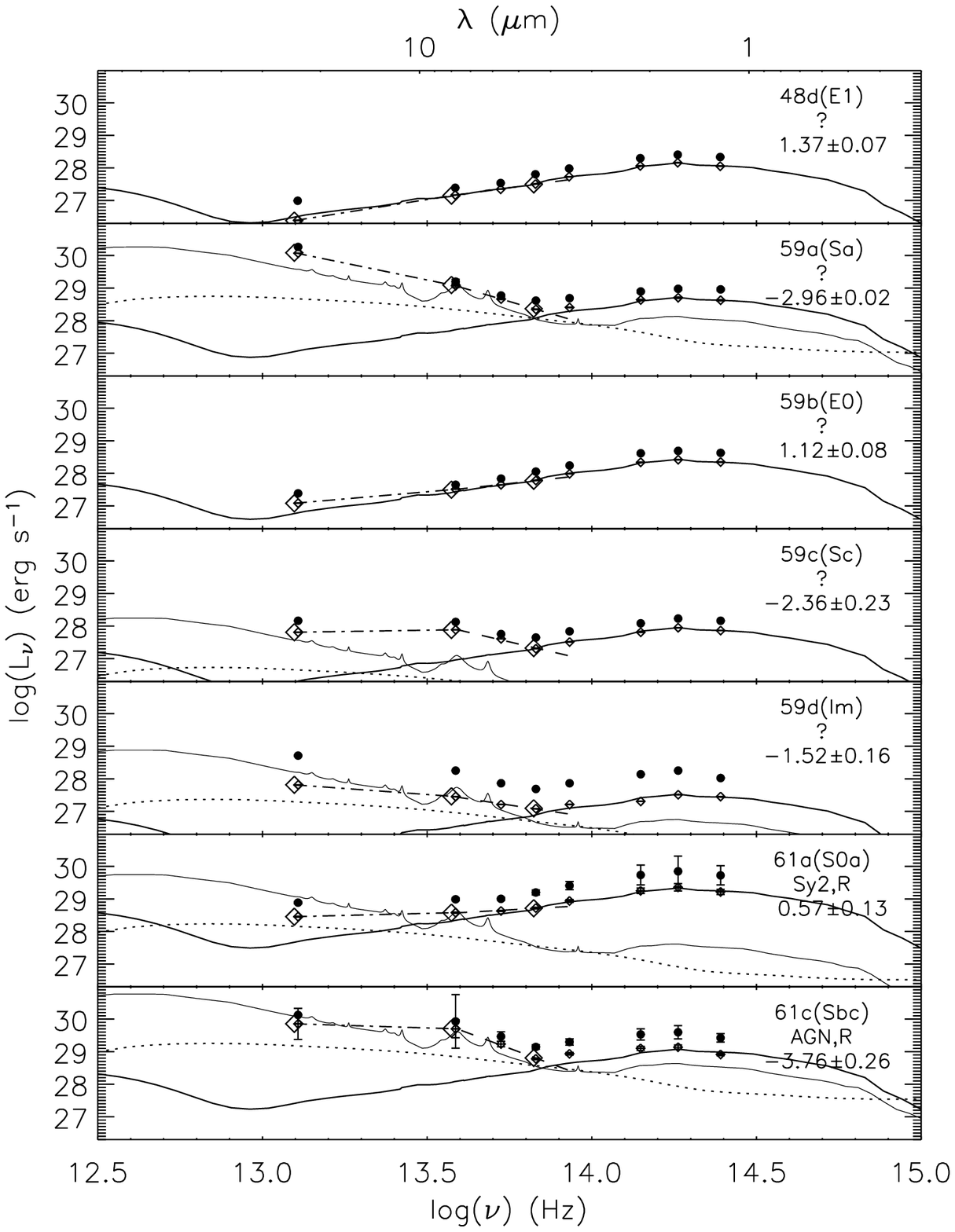}}\\
\centerline{Fig. 1. --- Continued.}
\clearpage
{\plotone{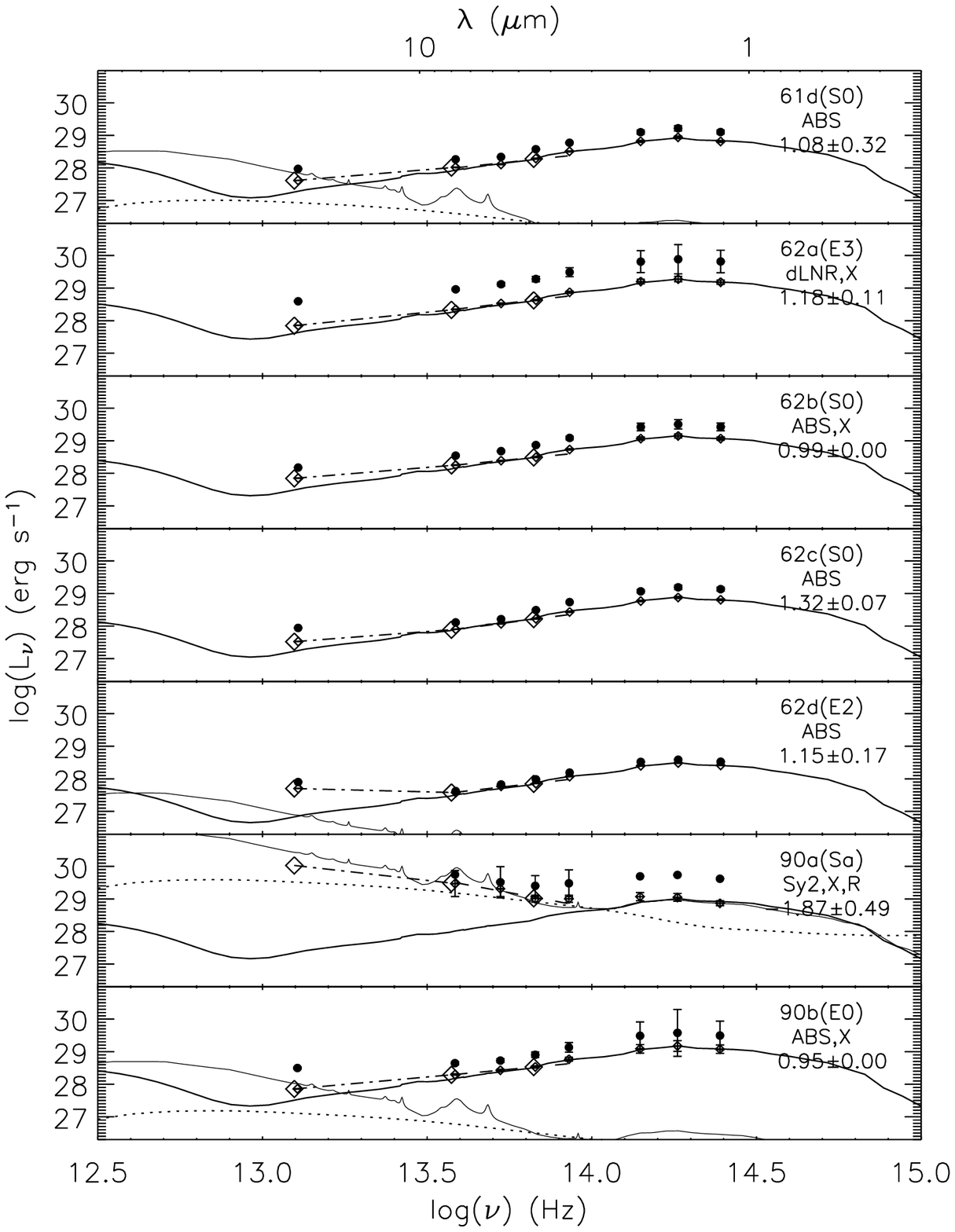}}\\
\centerline{Fig. 1. --- Continued.}
\clearpage
{\plotone{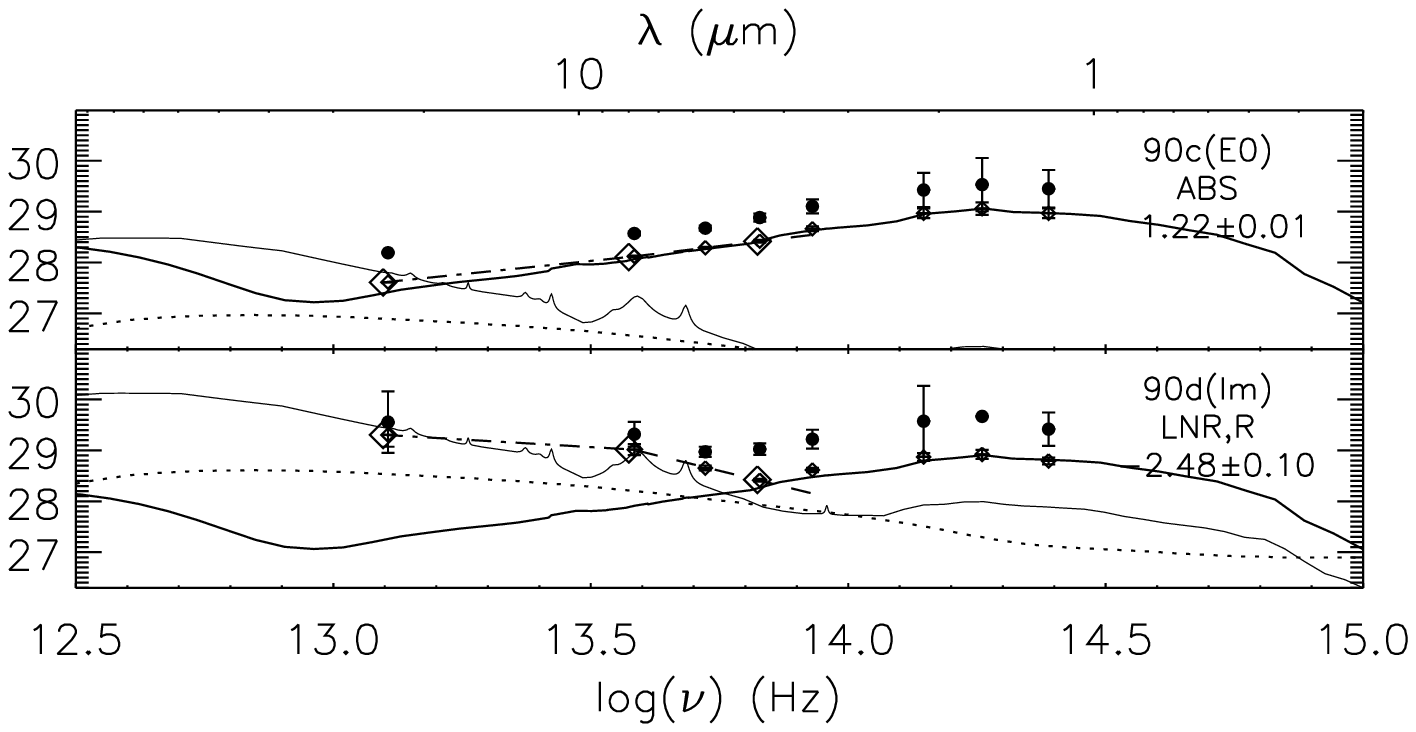}}\\
\centerline{Fig. 1. --- Continued.}
\clearpage

\figurenum{2}
\figurenum{3}
\figurenum{4}
\figurenum{5}
\clearpage
\figurenum{6}
\figurenum{7}

\end{document}